%% file: almost1.tex
\documentclass[12pt, leqno, notitlepage]{article}

\usepackage{thm-restate}

\usepackage{float}
\usepackage{latexsym,graphicx,amsmath,amssymb, amsthm, chngcntr, bm}
\usepackage{thmtools} 
\usepackage[usenames, dvipsnames]{color}
\usepackage[utf8]{inputenc}
\usepackage[numbers]{natbib}
\usepackage{xspace}
\usepackage{accents}
\usepackage{mleftright}%
\usepackage[cm]{fullpage}%
\usepackage{mleftright}%
\usepackage{comment}
\usepackage{mathrsfs} 
\usepackage{subcaption}
\usepackage{enumitem}
\usepackage{mathtools}
\usepackage{microtype}
\usepackage{euscript}
\usepackage{titlesec}
\usepackage{bold-extra}
\usepackage{etoolbox}
\usepackage{pdfpages}

\usepackage{pgf,tikz,pgfplots}
\pgfplotsset{compat=1.14}
\usepackage{mathrsfs}
\usetikzlibrary{arrows}

\titleformat{\section}
  {\Large \bfseries \sffamily}{\thesection}{1em}{}
\titleformat{\paragraph}
  {\normalfont\normalsize\bfseries}{\theparagraph}{1em}{}
\titlespacing*{\paragraph}
  {0pt}{3.25ex plus 1ex minus .2ex}{1.5ex plus .2ex}
\titlespacing{\subparagraph}{0pt}{10pt}{8pt}

\def\szemeredi{Szemer{\'e}di\xspace}

\def\reals{\mathbb{R}\xspace}

\def\sph{{\mathbb S}}

\def\W{{\cal W}}
\def\C{{\cal C}}
\def\L{{\cal L}}
\def\P{{\cal P}}

\def\sph{{\mathbb S}}

\def\W{{\cal W}}
\def\H{{\cal H}}

\def\set#1#2{{\lbrace #1 \mid #2 \rbrace}}

\newcommand{\norm}[1]{\left\lVert#1\right\rVert}

\newtheorem{theorem}{Theorem}[section]
\newtheorem*{remark*}{Remark}
\newtheorem{lemma}[theorem]{Lemma}

\begin{document}

\title{Incidences between points and curves with almost two degrees of freedom\thanks{%
Partially supported by ISF Grant 260/18, by grant 1367/2016
   from the German-Israeli Science Foundation (GIF), and by
   Blavatnik Research Fund in Computer Science at Tel Aviv University. A preliminary version
   of this paper has appeared in the Proceedings of the 36th Symposium on Computational Geometry, 2020.}
\author{Micha Sharir\thanks{%
School of Computer Science, Tel Aviv University, Tel~Aviv, Israel; {\tt michas@tau.ac.il}}
\and
Noam Solomon\thanks{%
School of Computer Science, Tel Aviv University, Tel~Aviv, Israel; {\tt noam.solom@gmail.com}}
\and
Oleg Zlydenko\thanks{%
School of Computer Science, Tel Aviv University, Tel~Aviv, Israel; {\tt zlydenko@gmail.com}}
}}
\maketitle

\begin{abstract}
We study incidences between points and (constant-degree algebraic) curves
in three dimensions, taken from a family $\C$ of curves that have 
\emph{almost two degrees of freedom}, meaning that (i) every pair of curves
of $\C$ intersect in $O(1)$ points, (ii) for any pair of points $p$, $q$,
there are only $O(1)$ curves of $\C$ that pass through both points, and
(iii) There exists a $6$-variate real polynomial $F$ of constant degree, so that
a pair $p$, $q$ of points admit a curve of $\C$ that passes through
both of them if and only if $F(p,q)=0$.
(As an example, the family of unit circles in $\reals^3$ that pass through 
some fixed point is such a family.)

We begin by studying two specific instances of this scenario. The first instance deals
with the case of unit circles in $\reals^3$ that pass through some fixed point
(so called \emph{anchored} unit circles).
In the second case we consider tangencies between \emph{directed points} and circles
in the plane, where a directed point is a pair $(p,u)$, where $p$ is a point in
the plane and $u$ is a direction, and $(p,u)$ is tangent to a circle $\gamma$
if $p\in\gamma$ and $u$ is the direction of the tangent to $\gamma$ at $p$.
A lifting transformation due to \citet{ESZ} maps these
tangencies to incidences between points and curves (`lifted circles') 
in three dimensions. In both instances we have a family of curves in
$\reals^3$ with almost two degrees of freedom.

We show that the number of incidences between $m$ points and $n$ anchored 
unit circles in $\reals^3$, as well as the number of tangencies between $m$ 
directed points and $n$ arbitrary circles in the plane, is
$O(m^{3/5}n^{3/5}+m+n)$ in both cases.

We then derive a similar incidence bound, with a few additional terms,
for more general families of curves in $\reals^3$
with almost two degrees of freedom, under a few additional natural assumptions.

The proofs follow standard techniques, based on polynomial partitioning, but
they face a critical novel issue involving the analysis of surfaces that are infinitely
ruled by the respective family of curves, as well as of surfaces in a dual
three-dimensional space that are infinitely ruled by the respective family of 
suitably defined dual curves.
We either show that no such surfaces exist, or develop and adapt techniques for handling
incidences on such surfaces.

The general bound that we obtain is $O(m^{3/5}n^{3/5}+m+n)$ plus additional terms 
that depend on how many curves or dual curves can lie on an infinitely-ruled surface.
\end{abstract}


\section{Introduction}

\subparagraph*{Our results: An overview.} 

In this paper we study several incidence problems involving points and algebraic curves
in three dimensions, where the curves are 3-parameterizable (each of them can be
defined by three real parameters) and have 
\emph{almost two degrees of freedom}, a notion that we discuss in detail below.
We begin by deriving improved incidence bounds for two specific classes of such
curves, one of which (studied in Section~\ref{ch:unit_circles})
is the class of \emph{anchored unit circles} (unit circles that pass 
through some fixed point), and the other (studied in Section~\ref{ch:point_circle})
is a class of `lifted circles'
that arise in the context of tangencies between so-called \emph{directed points}
and circles in the plane. In both cases, the incidence bound, for $m$ points and $n$ 
curves, is $O(m^{3/5}n^{3/5} + m + n)$. We then study the problem for general curves
that satisfy the above properties (and a few other natural assumptions), 
and derive the same bound as above, with additional terms that
depend on various parameters
associated with the problem. See Section~\ref{ch:generalizations} for full details. 

We begin with a review of the setup and of several basic features that 
arise in the analysis.

\subparagraph*{Incidence problems.} \label{sec:intro-definitions}

Let $P$ be a set of $m$ points, and let $C$ be a set of $n$ algebraic curves 
in $\reals^3$, taken from some inifnite family $\C$ of algebraic curves of 
some bounded degree (such as lines, circles, etc.). Let $I(P, C)$ denote 
the number of incidences between the points of $P$ and the curves of $C$, i.e., 
${\displaystyle I(P, C) = |\set{(p, c)}{p \in P, c \in C, p \in c}|}$.
The \emph{incidence problem} for $\C$ is to estimate
$I(m,n) \coloneqq \max_{|P| = m, |C| = n}{I(P ,C)}$, where the maximum is over 
all sets $P$ of $m$ points and $C$ of $n$ curves from $\C$.

The simplest formulation of the incidence problem
involves incidences between points and lines in the plane, where we have
\begin{theorem}[{\citet{SzemerediTrotter}}]  \label{thm:szemeredi_trotter}
  For sets $P$ of $m$ points and $L$ of $n$ lines in the plane, we have
  ${\displaystyle I(P, L) = O(m^{2/3}n^{2/3} + m + n)}$,
  and the bound is tight in the worst case.
\end{theorem}

The same asymptotic upper bound can be proven for unit circles as well, except 
that the matching lower bound is not known to hold, and is strongly suspected 
to be only close to linear. For general circles, of arbitrary radii, we have
\begin{theorem}[\citet{Agarwal} and \citet{MarcusTardos}] \label{thm:arbitrary_circles_plane}
  For sets $P$ of $m$ points and $C$ of $n$ (arbitrary) circles in the plane we
  have
  ${\displaystyle I(P, C) = O(m^{2/3}n^{2/3} + m^{6/11}n^{9/11}\log^{2/11}(m^3/n) + m + n)}$.
\end{theorem}

Another variant of the incidence problem, which has recently been studied in~\citet{ESZ}, 
(see also \citet{Zahl} for a related technique in three dimensions)
and which is relevant to the study in this paper, is bounding the number
of tangencies between lines and circles in the plane. In more detail, let a
\emph{directed point} in the plane be a pair $(p, u)$, where $p \in \reals^2$ 
and $u$ is a direction (parameterized by its slope). 
A tangency occurs between a circle $c$ and a directed 
point $(p, u)$ when $p \in c$ and $u$ is the direction of the tangent to $c$ at
$p$; see Figure~\ref{fig:directed_points_intro}. Unlike the standard case of 
point-circle incidences, there can be at most one circle that is tangent to a 
given pair of directed points (and in general there is no such circle). 
\citet{ESZ} showed:

\begin{theorem}[\citet{ESZ}] \label{thm:ellenberg}
  For a set $P$ of $m$ directed points and a set $C$ of $n$ (arbitrary) circles 
  in the plane, there are $O(n^{3/2})$ tangencies between the circles in $C$ 
  and the directed points in $P$, assuming that each point of $P$ is incident
  (i.e., tangent) to at least two circles. 
\end{theorem}

In fact, the bound in \cite{ESZ} also holds for more general sets of curves, and 
over fields other than $\reals$. An immediate corollary of Theorem~\ref{thm:ellenberg} 
is that the number of incidences between $m$ directed 
points and $n$ circles is $O(n^{3/2} + m)$. We will discuss this problem further 
in Section~\ref{ch:point_circle}, where we obtain the improved bound
$O(m^{3/5}n^{3/5}+m+n)$ mentioned above. 

\begin{figure}[htb]
    \centering
    \begin{minipage}{0.45\textwidth}
        \centering
        \resizebox{6cm}{!}{\input{figures/directed_points_intro.tikz}}
        \caption{ \small \sf The tangent to circle $c$ at point $p$ has direction
        $u$. We then say that the directed point $(p, u)$ is tangent (or incident) 
        to $c$. } 
        \label{fig:directed_points_intro}
    \end{minipage}\hfill
    \begin{minipage}{0.45\textwidth}
        \centering
        \resizebox{6cm}{!}{\input{figures/anchored_circles.tikz}}
        \caption{ \small \sf An anchored circle $c$. Its center is on the unit 
        sphere $\sph(o, 1)$, and it passes through the origin $o$. } 
        \label{fig:anchored_circles}
    \end{minipage}
\end{figure}

As has been observed, time and again, the result of 
Theorem~\ref{thm:szemeredi_trotter}, including both the upper and the lower bound,
is applicable to point-line incidences 
$\reals^3$ as well (and, in fact, in any higher-dimensional space $\reals^d$), unless we impose
some additional constraint on the number of coplanar input lines.
The following celebrated theorem of \citet{GuthKatz} gives such an 
improved bound\footnote{
  The theorem is not stated explicitly in \cite{GuthKatz}, but it is an 
  immediate consequence of the analysis in  \cite{GuthKatz}.
}.

\begin{theorem}[{\citet{GuthKatz}}]  \label{thm:guth_katz}
  Let $P$ be a set of $m$ points and $L$ be a set of $n$ lines in $\reals^3$. 
  Assume further that no plane in $\reals^3$ contains more than $q$ lines of
  $L$, for some parameter $q \le n$. Then
  ${\displaystyle I(P, L) = O\left( m^{1/2}n^{3/4} + m^{2/3}n^{1/3}q^{1/3} + m + n \right)}$.
  Moreover, the bound is tight in the worst case.
\end{theorem}

A similar argument can be made for point-circle incidences in $\reals^3$ 
(or again in any dimension $\ge 3$)---here we need to constrain the number
of input circles that can lie in any common plane or sphere.
The best known upper bounds, due to~\citet{SharirSolomon} and to~\citet{Zahl}, are 
(see also \citet{SSZ} for an earlier, weaker bound):
\begin{theorem}[\citet{SharirSolomon}] \label{thm:arbitrary_circles_space}
  Let $P$ be a set of $m$ points and let $C$ be a set of $n$ circles in
  $\reals^3$, and let $q < n$ be an integer. If no sphere or plane contains 
  more than $q$ circles of $C$, then
  \[
  I(P, C) = O\left( m^{3/7}n^{6/7} + m^{2/3}n^{1/3}q^{1/3} + 
  m^{6/11}n^{5/11}q^{4/11}\log^{2/11}(m^3/q) + m + n \right) .
  \]
\end{theorem}

\begin{theorem}[\citet{Zahl}] \label{thm:arbitrary_circles_space_1}
  Let $P$, $C$, $m$, $n$, and $q$ be as above. If no sphere or plane contains 
  more than $q$ circles of $C$, then
  \[
  I(P, C) = O^*\left( m^{1/2}n^{3/4} + m^{2/3}n^{13/15} + m^{1/3}n^{8/9}
  + nq^{2/3} + m \right) 
  \]
(where $O^*(\cdot)$ hides small sub-polynomial factors). 
\end{theorem}

\subparagraph*{Polynomial partitioning.} \label{subsec:intro-partitioning}

The polynomial partitioning technique is a powerful method 
for deriving incidence bounds (and many other results too), introduced 10 years ago
by~\citet{GuthKatz}, with an extended stronger version given later 
by~\citet{Guth}. We use the following version (specialized to our needs), where 
$Z(f)$ denotes the zero set $\set{z \in \reals^3}{f(z) = 0}$ of a real 
(trivariate) polynomial $f$.
\begin{theorem}[Polynomial partitioning \cite{Guth, GuthKatz}]
\label{thm:polynomial_part}
  Let $P$ be a set of $m$ points and $C$ be a set of $n$ algebraic curves of 
  some constant degree in $\reals^3$. Then, for any $1 < D$ such that 
  $D^3 < m$ and $D^2 < n$, there is a polynomial $f$ of degree at most $D$ 
  such that each of the $O(D^3)$ (open) connected components of 
  $\reals^3 \setminus Z(f)$ contains at most $O(m / D^3)$ points of $P$, 
  and is crossed by at most $O(n / D^2)$ curves of $C$. 
\end{theorem}
Note that the theorem has no guarantee regarding the number of points of $P$ on
$Z(f)$, or the number of curves of $C$ that are contained in $Z(f)$. 

One of the main techniques for proving incidence bounds via polynomial partitioning 
proceeds as follows. We first establish a simple (and weak) incidence bound 
(usually referred to as a \emph{bootstrapping} bound) by some other method. 
Then we apply Theorem~\ref{thm:polynomial_part}, use the bootstrapping bound 
in every connected component (cell) of $\reals^3 \setminus Z(f)$, and sum up
these bounds to obtain a bound on the number of incidences within the cells. 
Incidences between curves in $C$ and points on $Z(f)$ must be treated separately, using a 
different set of tools and techniques, typically taken from algebraic geometry.

\subparagraph*{Degrees of freedom.} \label{sec:intro-dof}

We say that a family $\C$ of constant-degree irreducible algebraic curves in
$\reals^3$ has \emph{$s$ degrees of freedom} (of multiplicity $\mu$) if: 
\begin{enumerate}
  \item each pair of curves of $\C$ intersect in at most $\mu$ points; and
  \item for each $s$-tuple $p_1, \ldots, p_s$ of distinct points in $\reals^3$ 
  there are at most $\mu$ curves of $\C$ that pass through all these points.
\end{enumerate}

The definition extends, verbatim, to curves in any other dimension.

The notion of degrees of freedom can be defined for arbitrary families of curves
(not necessarily algebraic). However, for various technical reasons, mainly 
to be able to apply Theorem~\ref{thm:polynomial_part}, we confine 
ourselves to the case of constant-degree algebraic curves.

Many natural families of curves have a small number of degrees of freedom:
\begin{itemize}
  \item Lines have two degrees of freedom with multiplicity one (in any space 
  $\reals ^ d$). Indeed, each pair of lines intersect in at most one point, and 
  through any pair of points only a single line can be drawn.
  \item Similarly, unit circles in the plane have two degrees of freedom as well, with 
  multiplicity two. 
  (Note that unit circles in $\reals ^ 3$, or in any higher-dimensional 
  space, do not have two degrees of freedom, but they have three degrees of 
  freedom, as a special case of the next example.)
  \item Circles of arbitrary radii, in any space $\reals^d$, have 
  three degrees of freedom.
\end{itemize}

The following theorem is a generalization of Theorem~\ref{thm:szemeredi_trotter}, 
and is due to \citet{PachSharir}. The original bound applies to more 
general families of curves, but we stick to the algebraic setup.
\begin{theorem}[\citet{PachSharir}] \label{thm:dof}
  Let $P$ be a set of $m$ points in the plane, and let $C$ be a set of $n$ 
  irreducible algebraic curves in the plane of degree at most $k$ and with $s$ 
  degrees of freedom (with multiplicity $\mu$); here $k$, $s$ and $\mu$ are 
  assumed to be constants. Then:
  \[
  I(P, C) = O\left( m^{\frac{s}{2s - 1}}n^{\frac{2s - 2}{2s - 1}} + m + n \right),
  \]
  where the constant of proportionality depends on $k$, $s$ and $\mu$.
\end{theorem}
Note that this bound coincides with the \szemeredi-Trotter bound for lines (for which $s=2$), and 
also with the bound for unit circles in the plane.

\begin{remark*}
  If we apply Theorem~\ref{thm:dof} to the family of circles of arbitrary radii, 
  in any dimension (for which $s = 3$), we get the bound
  $I(P, C) = O(m^{3/5}n^{4/5} + m + n)$, which is weaker than the bound in
  Theorem~\ref{thm:arbitrary_circles_plane}.
\end{remark*}

\subparagraph*{Infinitely ruled surfaces.} \label{sec:intro-infinitely_ruled}

Extending the constraint that the parameter $q$ imposes in Theorem~\ref{thm:guth_katz},
we use the following concept, studied by Sharir and Solomon in \cite{SharirSolomon},
building on a similar concept from \citet{GuthZahl}. An algebraic surface 
$V$ in $\reals^3$ is \emph{infinitely ruled} by a family $\C$ of curves, 
if each point $q \in V$ is incident to infinitely many curves of $\C$ that 
are fully contained in $V$. For example, the only surfaces that are infinitely 
ruled by lines are planes, and the only surfaces that are infinitely ruled by 
circles are planes and spheres; see \citet{Lubbes}. Sharir and Solomon have
considered this notion in \cite{SharirSolomon} to show:
\begin{theorem}[\citet{SharirSolomon}] \label{thm:dof_3d}
  Let $P$ be a set of $m$ points and $C$ a set of $n$ irreducible algebraic 
  curves in $\reals ^ 3$, taken from a family $\C$, so that the curves of $\C$ 
  are algebraic of constant degree,\footnote{%
    The analysis in \cite{SharirSolomon} also applies to \emph{constructible}
    families of curves, which generalizes the notion of algebraic curves.}
  and with $s$ degrees of freedom (of 
  some multiplicity $\mu$). If no surface that is infinitely ruled by curves of
  $\C$ contains more than $q$ curves of $C$, for a parameter $q < n$, then
  ${\displaystyle I(P,C) = O\left( m^{\frac{s}{3s - 2}}n^{\frac{3s - 3}{3s - 2}} + 
  m^{\frac{s}{2s - 1}}n^{\frac{s - 1}{2s - 1}}q^{\frac{s - 1}{2s - 1}} + m + n \right)}$,
  where the constant of proportionality depends on $s$, $\mu$, and the degree of the curves in $\C$.
\end{theorem}

Note that Theorem~\ref{thm:guth_katz} is a special case of this result, 
with $s = 2$, where the infinitely ruled surfaces are planes.

An additional tool that we rely on is also due to \citet{SharirSolomon}. It is 
the following theorem, which is part of Theorem 1.13 in \cite{SharirSolomon}, 
and is a generalization of a result of \citet{GuthZahl} (that was stated there
only for doubly ruled surfaces).

\begin{theorem}[\citet{SharirSolomon}] \label{thm:inf_ruled_exceptional}
  Let $\C$ be a family of algebraic curves in $\reals^3$ of constant degree $E$. 
  Let $f$ be a complex irreducible polynomial of degree $D \gg E$. If $Z(f)$ is not 
  infinitely ruled by curves from $\C$ then there exist absolute constants
  $c$, $t$, such that, except for at most $c D^2$ exceptional curves, every 
  curve in $\C$ that is fully contained in $Z(f)$ is incident to at most $c D$ 
  $t$-rich points, namely points that are incident to at least $t$ curves in $\C$ 
  that are also fully contained in $Z(f)$.
\end{theorem}

\subparagraph*{Almost two degrees of freedom.}

We introduce the following notion. A family $\C$ of algebraic irreducible curves 
in $\reals^3$ has \emph{almost $s$ degrees of freedom} (of multiplicity $\mu$) if: 

\medskip
\begin{enumerate}
  \item each pair of curves of $\C$ intersect in at most $\mu$ points;
  \item for each $s$-tuple $p_1, \ldots, p_s$ of distinct points in $\reals ^ 3$ 
  there are at most $\mu$ curves of $\C$ that pass through all these points; and
  \item there exists a curve of $\C$ that passes through $p_1, \ldots, p_s$, if 
  and only if $F(p_1, \ldots, p_s) = 0$, where $F$ is some $3s$-variate real 
  polynomial of constant degree associated with $\C$. 
\end{enumerate}
\medskip

With this definition we want to capture families $\C$ of curves that have some
$s$ degrees of freedom, but are such that for most $s$-tuples of points there is 
no curve of $\C$ that passes through all of them. As we demonstrate in this work,
this additional restriction helps us improve the upper bound for incidences 
between points and curves from such a family.

As with the case of standard degrees of freedom, there are natural examples that 
fall under this definition. One such example is the family of unit circles in
$\reals^3$ (or in any $\reals^d$, for $d\ge 3$), which, as is easily checked,
has almost three degrees of freedom, with multiplicity two. 

\subparagraph*{Our results.} \label{sec:intro-results}

Although the above definition applies for general values of $s$ and $d$, in this
paper we focus on the special case $s = 2$ and $d = 3$.

In Section~\ref{ch:unit_circles}, we study the incidence problem between points
and unit circles in three dimensions that pass through a fixed point (so-called
\emph{anchored} unit circles). With this additional constraint, this family has 
almost two degrees of freedom. We use this property to prove the bootstrapping 
bound $I(m, n) = O(m^{3/2} + n)$, which improves the naive bootstrapping bound 
$I(m, n) = O(m^2 + n)$ for general families of curves with two degrees of freedom. 
We then prove that no surface is infinitely ruled by this family of curves. 
Combining this with some additional arguments, most notably an argument that
establishes the absence of infinitely ruled surfaces in a suitably defined dual 
context (needed to establish our improved bootstrapping bound), gives us the 
following incidence bound:
\[
I(m, n) = O(m^{3/5}n^{3/5} + m + n) .
\]
We remark that Sharir et al.~\cite{SSZ} have obtained the bound
\begin{equation} \label{eq:ssz:unit}
I(m,n) = O^*(m^{5/11}n^{9/11} + m^{2/3}n^{1/2}q^{1/6} + m + n)
\end{equation}
for $m$ points and $n$ \emph{non-anchored} unit circles in $\reals^3$ 
(where, as above, $O^*(\cdot)$ hides small sub-polynomial factors). While this bound
applies to general families of unit circles, it does not imply our bound
for anchored circles (and it depends on the threshold parameter $q$, of
which our bound is independent).

In Section~\ref{ch:point_circle}, we bound the number of tangencies between 
circles and directed points in the plane. We transform this problem to 
an incidence problem between points and curves with almost two degrees of 
freedom in $\reals^3$, resulting from lifting the given 
circles to three dimensions, using a method of \citet{ESZ}. In this case as 
well, we prove the bootstrapping bound\footnote{%
  Note the difference between this bound and the bound in Ellenberg et al.~\cite{ESZ} 
  noted earlier. It is this stronger version that allows us to derive our bound,
  mentioned below.}
$I(m, n) = O(m^{3/2} + n)$, 
show that no surface is infinitely ruled by this family of curves, and combine 
these statements (with some other considerations) to get the same asymptotic 
bound $I(m, n) = O(m^{3/5}n^{3/5} + m + n)$. 
This bound is stronger than the bound $O(n^{3/2})$ (or rather $O(n^{3/2}+m)$)
derived in \cite{ESZ}.

In Section~\ref{ch:generalizations}, we extend the 
proofs from Sections~\ref{ch:unit_circles} and \ref{ch:point_circle}, for more 
general families of curves with almost two degrees of freedom in three 
dimensions. A large part of the analysis can be generalized directly, but
in general, there may exist surfaces that are infinitely ruled by
these families of curves. Additionally, as already noted, our analysis in 
Sections~\ref{ch:unit_circles} and \ref{ch:point_circle} also involves a stage 
where it studies the problem in a dual setting, and the existence of
infinitely ruled surfaces is an issue that has to be dealt with in this setting too.
As in Theorem~\ref{thm:guth_katz}, the bound depends on the maximum number 
of curves that can lie on a surface that is infinitely ruled by the given 
family of curves, and on a similar threshold parameter in the dual space. 
We also need to impose a few additional natural conditions on the family of curves 
to obtain our result.

The bound that we obtain is $O(m^{3/5}n^{3/5} + m + n)$ plus additional terms 
that depend on the threshold parameters for infinitely ruled surfaces, both in 
the primal and in the dual setups. These terms are subsumed in the bound just 
stated when the relevant parameters are sufficiently small. See
Section~\ref{ch:generalizations} for the precise bound.

We exemplify the general bound for families of lines in $\reals^3$ that
have almost two degrees of freedom, a problem that has also been looked at
by Guth et al. (work in progress).

We conclude the paper in Section~\ref{ch:conclusion} by listing some open 
problems and suggesting directions for further research.


\section{Anchored unit circles in space} \label{ch:unit_circles}

\subparagraph*{The setup.}
As stated in Section~\ref{sec:intro-results}, unit circles in space have almost 
three degrees of freedom. We reduce the setup to one with almost two degrees of 
freedom, by considering only circles that pass through a fixed point, say the 
origin. We call such circles \emph{anchored (unit) circles}.
An anchored circle $c$ has radius $1$ and passes through the origin, so
its center lies on the unit sphere $\sph(o, 1)$ centered at $o$ 
(see Figure~\ref{fig:anchored_circles}). The main result of this section is

\begin{restatable}{theorem}{thmanchoredmain}\label{thm:anchored_main}
  The number of incidences between $m$ points and $n$ anchored circles in
  $\reals^3$ is \[I(P, C) = O(m^{3/5}n^{3/5} + m + n).\]
\end{restatable}

\subsection{Proof of Theorem~\ref{thm:anchored_main}}\label{sec:anchored_thm_proof}

We obtain the desired bound by following the general approach in
\cite{SharirSolomon}. Using special properties of the underlying setup, 
we obtain the following improved bootstrapping bound (over the 
simple ``naive'' bound $O(m^2 + n)$ used in \cite{SharirSolomon}).
\begin{restatable}{lemma}{lemanchoredboot}\label{lem:anchored_boot}
  The number of incidences between a set $P$ of $m$ points and a set $C$ of $n$ 
  anchored unit circles in $\reals^3$ is $I(P, C) = O(m^{3/2} + n)$.
\end{restatable}

The proof of the lemma is given in Section~\ref{sec:anchored_boot_proof} below. 
Assuming for now that the lemma holds, we apply the technique of
\cite{SharirSolomon}, with suitable modifications, to derive the incidence bound 
in Theorem~\ref{thm:anchored_main}. We show, by induction on $n$, that
$I(P,C) \le A\left(m^{3/5}n^{3/5} + m + n\right)$, for a suitable constant $A$, 
to be chosen later.
It is trivial to verify that this bound holds for $n$ smaller than some constant
threshold $n_0$, so we focus on the induction step.

We first construct, using Theorem~\ref{thm:polynomial_part}, a partitioning 
polynomial $f$ in $\reals^3$, of some specified (maximum) degree $D$, so that 
each cell (connected component) of $\reals^3 \setminus Z(f)$ contains at most 
$O(m / D^3)$ points of $P$, and is crossed by at most $O(n / D^2)$ circles of $C$.

For each (open) cell $\tau$ of the partition, let $P_\tau$ denote the set of 
points of $P$ inside $\tau$, and let $C_\tau$ denote the set of circles of $C$ 
that cross $\tau$; we have $m_\tau := |P_\tau| = O(m / D^3)$, 
and $n_\tau := |C_\tau| = O(n / D^2)$. We apply the bootstrapping bound of
Lemma~\ref{lem:anchored_boot} within each cell $\tau$, to obtain
\[
I(P_\tau,C_\tau) = 
O\left( m_\tau^{3/2} + n_\tau \right) =
O\left( (m / D^3)^{3/2} + (n / D^2) \right) = 
O\left( m^{3/2} / D^{9/2} + n / D^2 \right) .
\]
Multiplying by the number of cells, we get that the number of incidences within 
the cells is
\[
  \sum_\tau I(P_\tau,C_\tau) = 
  O\left( D^3\cdot \left( m^{3/2}/D^{9/2} + n/D^2 \right) \right) = 
  O\left( m^{3/2}/D^{3/2} + nD \right) . 
\]

We choose $D = am^{3/5}/n^{2/5}$, for a sufficiently small constant $a$. For 
this to make sense, we require that $1 \le D \le a' \min\{m^{1/3}, n^{1/2}\}$, 
for another sufficiently small constant $a' > 0$, which holds when
$b_1 n^{2/3} \le m \le b_2 n^{3/2}$, for suitable constants $b_1, b_2$ that 
depend on $a$ and $a'$. 

If $m < b_1 n^{2/3}$, the bound in Lemma~\ref{lem:anchored_boot} yields (for the 
entire sets $P$, $C$) the bound $O(m^{3/2} + n) = O(n)$. 

If $m > b_2 n^{3/2}$, we construct a partitioning polynomial $f$ of degree
$D = a' n^{1/2}$, for the sufficiently small constant $a'$ introduced above, so that each cell of
$\reals^3 \setminus Z(f)$ contains at most $O(m / D^3)$ points of $P$ and is 
crossed by at most $O(n / D^2) = O(1)$ circles of $C$. The number of incidences 
within each cell is then at most $O(m / D^{3})$, for a total of
$O(D^3) \cdot O(m / D^3) = O(m)$ incidences. More precisely, we write this bound
as $O(m_0)$, where $m_0$ is the number of points of $P$ within the cells. We 
also denote by $m^*$ the number of points of $P \cap Z(f)$, so $m_0 + m^* = m$. 
Handling incidences on the zero set $Z(f)$ is done as in the case of a smaller
$m$ (in the middle range), as detailed shortly.

To recap, if $m$ is in the middle range, we substitute $D = am^{3/5} / n^{2/5}$ 
and obtain the bound $O(m^{3/5}n^{3/5})$ for incidences within the cells. 
If $m$ is in the range of small values $m \le b_1 n^{2/3}$, no partitioning 
is needed, and the bound is only $O(n)$. Finally, as just noted, 
for $m$ in the range of large values, i.e., for
$m > b_2 n^{3/2}$, the contribution within the cells is $O(m)$.
(We note that, as follows from the analysis of \citet{ESZ}, if we only
consider directed points incident to at least two circles, then $m = O(n^{3/2})$.
A similar, somewhat weaker statement, also follows from 
the final bound that we will obtain.)
Altogether, the incidence bound within the cells is
\begin{equation} \label{eq:anchored_in_cells}
O\left( m^{3/5}n^{3/5} + m + n \right) .
\end{equation}

Consider next incidences involving points that lie on $Z(f)$. A circle $\gamma$ 
that is not fully contained in $Z(f)$ crosses it in at most $O(D)$ points, which
follows from B\'ezout's theorem (see, e.g., \cite{CLO}). This yields a total of
$O(nD) = O(m^{3/5}n^{3/5} + m)$ incidences (for $m = \Omega(n^{3/2})$, our choice 
of $D$ gives $O(nD) = O(n^{3/2} = O(m)$), within the asymptotic bound as in 
(\ref{eq:anchored_in_cells}). It therefore remains to bound the number of 
incidences between the points of $P$ on $Z(f)$ and the anchored circles that 
are fully contained in $Z(f)$. 

We follow the proof of Theorem 1.4~in \cite{SharirSolomon}, which considers each 
irreducible component of $Z(f)$ separately, and distinguishes between components
that are \emph{infinitely ruled} by anchored circles, and components that are not. 
Let $\C$ denote the infinite family of all possible anchored (unit) circles. 
Fortunately for us, we have:
\begin{restatable}{lemma}{lemanchorednoinfruled}\label{lem:anchored_no_inf_ruled}
  No algebraic surface is infinitely ruled by anchored unit circles.
\end{restatable}

\noindent{\bf Proof.}
Assume to the contrary that there exists an algebraic surface $V$ that is 
infinitely ruled by curves from $\C$; assume also, without loss of generality,\footnote{%
 Each $q\in V$ lies on at least one irreducible component that contains
 infinitely many curves of $\C$ through $q$, so at least one of the at most
 ${\rm deg}(f)$ components contains infinitely many points with this property, 
 and then one can show that this component is infinitely ruled.}
that $V$ is irreducible. It is known that the only irreducible surfaces that are
infinitely ruled by circles are spheres and planes (see, e.g., \cite{Lubbes}). 

Let then $V$ be a plane or a sphere. In order to contain any
$\gamma \in \C$, $V$ must pass through $o$. However, for any generic point $q\in V$,
there can be at most two unit circles that pass through $q$ and $o$ and are
contained in $V$, implying that $V$ cannot be infinitely ruled by such circles.
$\Box$\medskip

Write $m^* = |P\cap Z(f)|$ and $m_0 = |P\setminus Z(f)|$, so $m = m_0+m^*$. 
The analysis in \cite{SharirSolomon}, which we follow here, handles each
irreducible component of $Z(f)$ separately. Enumerate these components as
$Z(f_1), \ldots, Z(f_k)$, for suitable irreducible polynomials
$f_1, \ldots, f_k$, of respective degrees $D_1, \ldots, D_k$, where
$\sum_{i=1}^k D_i \le D$. By Lemma~\ref{lem:anchored_no_inf_ruled}, none of 
these components is infinitely ruled by anchored circles. 

Let $P_i$ (resp., $C_i$) denote the set of all points of $P$ (resp., anchored circles of $C$)
that are contained (resp., fully contained) in $Z(f_i)$, assigning 
each point and circle to the first such component (in the above order). 
The `cross-incidences', between points and circles 
assigned to different components, occur at crossing points between circles and 
components that do not contain them, and their number is easily seen to be $O(nD)$, 
which satisfies our asymptotic bound. It therefore suffices to bound the number of
incidences between points and circles assigned to the same component.

By Theorem~\ref{thm:inf_ruled_exceptional}, 
there exist absolute constants $c$, $t$, such that there are at most $c D_i^2$ 
`exceptional' anchored circles in $C_i$, namely, anchored circles that contain more 
than $c D_i$ \emph{$t$-rich} points of $P \cap Z(f_i)$, namely points
that are incident to at least $t$ circles from $C_i$. Denote the number of 
$t$-rich points (resp., \emph{$t$-poor} points, namely points that are not 
$t$-rich) as $m_{rich}$ (resp., $m_{poor}$), so
$m_{rich} + m_{poor} = m^*$. By choosing $a$ and $a'$ (in the definition of $D$) 
sufficiently small, we can ensure, as is easily checked, that 
$\sum_i D_i^2 \le (\sum_i D_i)^2 \le D^2 \le n/(2c)$. 

The number of incidences on the non-exceptional circles, summed over all 
components $Z(f_i)$, is $O(m_{poor} + nD)$. Indeed, each non-exceptional circle 
contains at most $c D_i$ $t$-rich points, for a total of $O(n D_i)$ incidences, 
and the sum of these bounds is $O(n D)$. Any $t$-poor point lies on at most $t$ 
circles of $C_i$, for a total of $t m_{poor} = O(m_{poor})$ incidences (over all 
sets $C_i$).

For the exceptional circles, we apply the induction hypothesis, as their overall 
number is at most $c \sum_i D_i^2 \le c D^2 \le n/2$. Note that in
this inductive step we only need to consider the $t$-rich points, as the
$t$-poor points have already been taken care of. By the induction hypothesis, 
the corresponding incidence bound between the points and circles that were 
assigned to (the same) $f_i$ is at most
\[
  A \left( m_i^{3/5}(c D_i^2)^{3/5} + m_i + c D_i^2 \right) ,
\]
where $m_i$ is the number of $t$-rich points assigned to $f_i$. We now sum over
$i$. Clearly, $\sum_i m_i = m_{rich}$. We also have $\sum_i c D_i^2 \le n / 2$. 
As for the first term, we use Hölder's inequality:
\small
\[
  \sum_i m_i^{3/5}(c D_i^2)^{3/5} = c^{3/5} \sum_i m_i^{3/5}D_i^{6/5} \le
  c^{3/5} \left( \sum_i m_i \right)^{3/5} \left( \sum_i D_i^3 \right)^{2/5} \le
  c^{3/5} m^{3/5} \left( \sum_i D_i^3 \right)^{2/5}.
\]
\normalsize
Finally, using the fact that $\sum_i D_i^3 \le D^3$, we get the overall bound:
\[
  A \left( c^{3/5}m^{3/5}D^{6/5} + m_{rich} + n / 2 \right) \le
  A \left( \frac{m^{3/5}n^{3/5}}{2^{3/5}} + m_{rich} + n / 2 \right) ,
\]
since $c^{3/5}D^{6/5} \le (n / 2)^{3/5}$, by construction.

We now add to this quantity the bound for incidences within the cells, as well as the 
various other bounds involving points on $Z(f)$. Together, we can upper bound these bounds by
$$
B \left( m^{3/5}n^{3/5} + n + m_0 + m_{poor} \right) ,
$$
for a suitable absolute constant $B$. By choosing $A$ sufficiently large, 
the sum of all the bounds encountered in the analysis is at most
${\displaystyle A \left( m^{3/5}n^{3/5} + m + n \right)}$. This establishes the induction
step, and thereby completes the proof of Theorem~\ref{thm:anchored_main}, modulo the still 
missing proof of Lemma~\ref{lem:anchored_boot}, presented next.
$\Box$\medskip

\subsection{Proof of Lemma~\ref{lem:anchored_boot}} \label{sec:anchored_boot_proof}

The lemma improves upon the naive (and standard) bootstrapping bound, used in
\cite{SharirSolomon}, which is $O(m^2 + n)$, for $m$ points and $n$ anchored circles. We
dualize the setup, exploiting the underlying geometry, mapping each circle $\gamma \in \C$ 
to a suitable algebraic representation of the point
$q_\gamma = (\alpha_\gamma, \beta_\gamma, \phi_\gamma)$ in 3-space, where
$(\alpha_\gamma, \beta_\gamma)$ represents the center of $\gamma$ as a point\footnote{%
  For example, we can use the $(x,y)$-coordinates of the center, applying a separate analysis
  to the upper hemisphere and the lower hemisphere, or we can use a standard algebraic
  re-parameterization of the spherical coordinates of the center.}
on $\sph(o, 1)$, and $\phi_\gamma$ represents the angle by which the circle is rotated around
the line connecting $o$ to its center. We denote by $\C^*$ the family of all these dual
points $q_\gamma$ (over all possible anchored unit circles $\gamma$). 

We also map each point $p \in P$ to the locus $h_p$ of all dual points $q_\gamma$ that 
represent anchored circles $\gamma$ that are incident to $p$, and argue 
that $h_p$ is a one-dimensional curve. To see this, we note that we may assume that 
$\|p\| \le 2$, for otherwise $p$ is not incident to any anchored unit circle. 
Assume first that $\|p\| < 2$. We first note that the
$(\alpha, \beta)$-projection of $h_p$ is a circle (on $\sph(o, 1)$). Indeed, if $\gamma$ is
anchored and incident to $p$, its center must lie on the intersection of the two spheres
$\sph(o, 1)$ and $\sph(p, 1)$, which is a circle (since $\|p\| < 2$). 
Hence we only have one degree of freedom
for choosing the $(\alpha, \beta)$-projection $(\alpha_\gamma, \beta_\gamma)$ of $q_\gamma$.
Moreover, after fixing $(\alpha_\gamma, \beta_\gamma)$, the rotation angle $\phi_\gamma$ is
also uniquely determined by the constraint that $p \in \gamma$. 
If $\|p\|=2$, the point $(\alpha_\gamma,\beta_\gamma)$ is unique (it is the midpoint of $op$),
but we still have one degree of freedom, or rotating $\gamma$ around $op$, so $h_p$
is a curve in this case too. Hence, in either case, $h_p$ is indeed a curve.

Let $\H$ denote the family of all dual curves $h_p$ for points $p \in \reals^3$ (actually, 
only points in the ball of radius $2$ around $o$, with $o$ excluded, are relevant). We show 
that $\H$ has (almost) two degrees of freedom. Clearly, all the curves in $\H$ are distinct,
and any pair of them intersect at most once (there is a unique circle that passes through 
the two corresponding primal points and $o$). Let $q_\gamma$, $q_{\gamma'}$ be
two distinct points of $\C^*$, representing two distinct anchored circles $\gamma, \gamma'$.
These circles intersect in at most two points, one of which is $o$, so that there is at most
one point $p \ne o$ that is incident to both $\gamma$ and $\gamma'$. That is, there is at
most one curve $h_p$ that passes through both $q_\gamma$ and $q_{\gamma'}$. (This argument
shows in fact that $\H$ has almost two degrees of freedom. That is, there is a polynomial
$G$ such that $G\left(q_\gamma,q_{\gamma'}\right) = 0$ if and only if $\gamma$ and $\gamma'$
intersect at a point other than $o$. However, we will not use this stronger property.)

Let $C^* \subset \C^*$ be the set of points $q_\gamma$ dual to the anchored 
circles $\gamma \in C$, and let $H \subset \H$ be the set of curves $h_p$ dual 
to the points $p \in P$. We have thus reduced our problem to that of bounding 
the number of incidences between $C^*$ and $H$, to which we can apply 
Theorem~\ref{thm:dof_3d}, using the fact that the curves of $\H$ have two 
degrees of freedom, to get the bound
\begin{equation} \label{eq:anchored_thm14_bound}
  I(P, C) = I(C^*, H) = 
  O\left( n^{1/2}m^{3/4} + n^{2/3}m^{1/3}q^{1/3} + n + m \right) , 
\end{equation}
where $q$ is the maximum number of curves from $H$ that lie on a common surface 
that is infinitely ruled by $\H$. Fortunately again for us, we have:

\begin{figure}[htb]
  \begin{center}
    \resizebox{6cm}{!}{\input{figures/infruled_lemma.tikz}}
    \caption{ \small \sf $\gamma$ is an anchored circle, $\gamma'$ is any 
    anchored circle with $q_{\gamma'}\in V$, and there are infinitely many such anchored circles; 
    by construction,
    it must intersect $\gamma$ 
    at some point $p$. For a generic choice of $\gamma'$ and of $r\in \gamma'$, all the anchored
    circles through $r$ must intersect $\gamma$, which is impossible.}
    \label{fig:infruled_lemma}
  \end{center}
\end{figure}
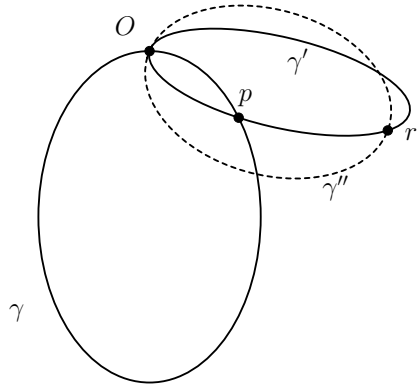

\begin{restatable}{lemma}{lemanchoreddualnoinfruled}\label{lem:anchored_dual_no_inf_ruled}
  No algebraic surface is infinitely ruled by $\H$.
\end{restatable}
\noindent{\bf Proof.}
Assume to the contrary that there exists an algebraic surface $V$ that is infinitely ruled 
by curves from $\H$; assume also, without loss of generality, that $V$ is irreducible (see a
previous footnote for the justification of this assumption), and write $V$ as the zero set
$Z(f)$ of an irreducible trivariate real polynomial $f$.

Fix a point $q_\gamma \in V$, and let $S(q_\gamma)$ denote the union of all dual curves
$h_p$ that pass through $q_\gamma$. By definition, this is equivalent to the condition that
$p \in \gamma\setminus\{o\}$, so $S(q_\gamma) = \bigcup_{p \in \gamma\setminus\{o\}} h_p$, 
which is easily seen to be a two-dimensional (dual) surface. 
Indeed, put $W = \{ (z,\tau) \mid \tau\in H,\;z\in \tau,\; q_\gamma \in\tau\}$.
Then $W$ is clearly an algebraic surface, and $S(q_\gamma)$ is the projection of $W$
onto the first coordinate $z$. (Compare with similar reasoning in \cite[Lemma 11.6]{GuthZahl}, 
and see also Sharir and Solomon~\cite[End of Section 3, p.~117]{SS:23}.)
By assumption, $S(q_\gamma) \cap V$ is also two-dimensional,
since the infinitely ruled surface $V$ contains infinitely many curves $h_p$ that are 
incident to $q_\gamma$ (and this infinite union of one-dimensional distinct curves must be
two-dimensional). Therefore, we must have $V \subseteq S(q_\gamma)$. 

Now take a point $q_{\gamma'} \ne q_\gamma \in V$. Then $q_{\gamma'} \in S(q_\gamma)$. This
means that there is a point $p \in \gamma$ (other than the origin) such that
$q_{\gamma'} \in h_p$. Back in the primal space, this implies that $\gamma$ intersects
$\gamma'$ at $p$. 

From the assumption on $V$, there are infinitely many curves $h_r$ that are fully contained
in $V$ and incident to $q_{\gamma'}$ (meaning, $r$ lies on $\gamma'$). Since $V\subseteq S(q_\gamma)$, 
any point $q_{\gamma''} \ne q_\gamma \in h_r$ (that is, any anchored circle $\gamma''$ through
$r$) must be contained in $S(q_\gamma)$, that is, must belong to some $h_s$ for
$s\in\gamma$. Back in the primal, this means that every anchored circle $\gamma''$ through
$r$ must intersect $\gamma$, which is clearly impossible for a generic choice of $r$; see
Figure~\ref{fig:infruled_lemma} for details. 
$\Box$\medskip

It thus follows that 
${\displaystyle I(P, C) = I(C^*, H) = O\left( n^{1/2}m^{3/4} + n + m \right)}$, which is 
upper bounded by ${\displaystyle O(n + m^{3/2})}$. This completes the proof of 
Lemma~\ref{lem:anchored_boot}, and, consequently, also of Theorem~\ref{thm:anchored_main}.
$\Box$\medskip


\section{Point-circle tangencies in the plane} \label{ch:point_circle}

\subparagraph*{The setup.}

Let $C$ be a set of $n$ circles in the plane. A \emph{directed point} in the plane is a pair
$(p, u)$, where $p \in \reals^2$ and $u$ is a direction. A circle $c$ is said to be
\emph{incident} (or \emph{tangent}) to a directed point $(p, u)$ if $c$ passes through $p$, 
and $c$ is tangent to the line emanating from $p$ in direction $u$. See 
Figure~\ref{fig:directed_points_intro}. 

As stated in Theorem~\ref{thm:ellenberg},
\citet{ESZ} (using a somewhat different notation) have shown that the number of directed 
points that are incident to at least two circles of $C$ is $O(n^{3/2})$. (Their bound also 
holds for more general families of algebraic curves.) Using the main technical idea in
\cite{ESZ}, we represent directions by their slopes\footnote{%
  This excludes $y$-vertical directions from the analysis. We assume, without loss of 
  generality, that no input directed point has vertical direction (i.e., slope
  $\pm \infty$).},
and regard each directed point $(p, u)$ as a point in $\reals^3$, where the $z$-coordinate 
is the slope; from now on, we let the parameter $u$ denote the slope. We map each circle $c$ 
in $\reals^2$ to the curve 
\[
c^* = \{(p, u) \mid \text{$c$ is incident to $(p, u)$} \} ,
\]
to which we refer as a \emph{lifted circle}, or the \emph{lifted image} of $c$. Note that,
since we represent directions by slopes, $c^*$ is an unbounded curve, consisting of two 
unbounded connected components. Moreover, as is easily checked, $c^*$ is an algebraic curve
of degree $4$. Specifically, a circle $c$ and a point $(p, u)$ that is incident to $c$ must 
satisfy the following two equations:
\[(p_x - c_x)^2 + (p_y - c_y)^2 = c_R^2 ,\quad\text{and}\quad 
u (p_y - c_y) = c_x - p_x ,\]
where $p = (p_x, p_y)$ and $c$ is a circle of radius $c_R$ centered at $(c_x, c_y)$. Hence, 
we can represent $c^*$ as the common zero set of the two polynomials
\begin{equation} \label{eq:lifted_circle}
  (x - c_x)^2 + (y - c_y)^2 = c_R^2 ,\quad\text{and}\quad 
  z (y - c_y) = c_x - x .
\end{equation}
As each of these polynomials is of degree $2$, $c^*$ is of degree $4$, as claimed.

\begin{figure}[htb]
  \begin{center}
    \resizebox{6cm}{!}{\input{figures/directed_points.tikz}}
    \caption{ \small \sf Let $(p, u)$ and $(q, v)$ be a pair of directed points. Let $w$ 
    denote the intersection of the perpendiculars to $(p, u)$ and $(q, v)$ through $p$ and
    $q$, respectively. A simple geometric argument shows that $(p, u)$ and $(q, v)$ have a 
    common incident (i.e., tangent) circle if and only if $|pw|=|qw|$. This can be easily 
    rewritten as $F(p, u, q, v) = 0$, for some fixed constant degree $6$-variate polynomial $F$.} 
    \label{fig:directed_points}
  \end{center}
\end{figure}
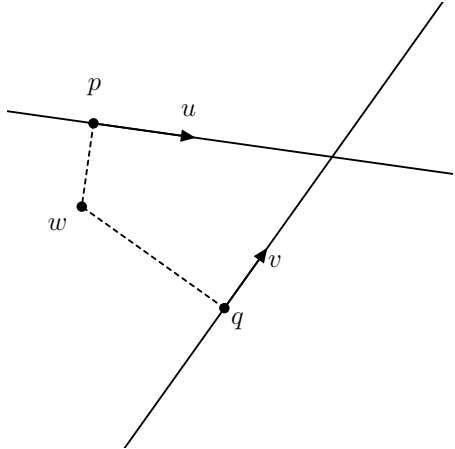

Denote by $\C$ the infinite family of all possible lifted circles. We claim that the curves
of $\C$ have almost two degrees of freedom. 
Indeed, two directed points define at most one circle that is incident to both of them; and
there exists a circle that is incident to two directed points $(p, u), (q, v)$ if and only 
if $(p, u, q, v)$ satisfies a (fixed) polynomial equation, as illustrated in 
Figure~\ref{fig:directed_points}.

The setup then becomes similar to what we have seen in Section~\ref{ch:unit_circles}, and we have 
\begin{restatable}{theorem}{thmtangentmain}\label{thm:tangent_main}
  The number of incidences between $m$ directed points and $n$ circles in the plane is 
  $O(m^{3/5}n^{3/5} + m + n)$.
\end{restatable}

\subsection{Proof of Theorem~\ref{thm:tangent_main}} \label{sec:tangent_thmp_roof}

The high-level approach in the proof of the theorem is very similar to the one presented in 
the previous section. Nevertheless, since the geometric details are different, and because 
of certain differences in the analysis, we spell out most of the details of the analysis, 
with some risk of redundancy.

Let then $P$ be a set of $m$ directed points and $C$ a set of $n$ circles in the plane. 
Write $C^* = \set{c^*}{c \in C}$, using the lifting defined earlier, and regard $P$ 
as a set of points in $\reals^3$. The goal is to estimate the number $I(P, C) = I(P, C^*)$ of 
incidences between $P$ (as a set of points in $3$-space) and $C^*$ (as a set of curves in
$3$-space).

We obtain the desired bound by following the same high-level approach as in 
Section~\ref{ch:unit_circles}. Using special properties of the underlying setup, we obtain 
the following improved bootstrapping bound (similar to the improved bootstrapping bound in 
Lemma~\ref{lem:anchored_boot}).
\begin{restatable}{lemma}{lemtangentboot}\label{lem:tangent_boot}
  The number of incidences between $m$ directed points and $n$ circles in the plane is
  $O(m^{3/2} + n)$.
\end{restatable}
The proof of the lemma is given in Section~\ref{sec:tangent_boot_proof} below. Assuming that 
the lemma holds, we prove, by induction on $n$ that, for $|P| = m$ and $|C| = n$,
$I(P, C) \le A (m^{3/5}n^{3/5} + m + n)$, for a suitable absolute constant $A$. Again, the
case of small $n$ is trivial, with a suitable choice of $A$, and we only focus on the induction step.

As before, we first construct a partitioning polynomial $f$ in $\reals^3$, of 
some specified (maximum) degree $D$, so that each cell (connected component) of
$\reals^3 \setminus Z(f)$ contains at most $O(m / D^3)$ points of $P$, and is 
crossed by at most $O(n / D^2)$ curves of $C^*$; the existence of such a 
polynomial follows from Theorem~\ref{thm:polynomial_part}.

For each (open) cell $\tau$ of the partition, let $P_\tau$ denote the set of 
points of $P$ inside $\tau$, and let $C^*_\tau$ denote the set of curves of
$C^*$ that cross $\tau$; we have
$m_\tau := |P_\tau| = O(m / D^3)$, and $n_\tau := |C^*_\tau| = O(n / D^2)$. We 
apply the bootstrapping bound of Lemma~\ref{lem:tangent_boot} within each cell
$\tau$, to obtain
\[
I(P_\tau, C^*_\tau) =
O\left( m_\tau^{3/2} + n_\tau \right) =
O\left( (m / D^3)^{3/2} + (n / D^2) \right) =
O\left( m^{3/2} / D^{9/2} + n / D^2 \right) .
\]
Multiplying by the number of cells, we get that the number of incidences within 
the cells is
\begin{equation} \label{eq:tangent_in_cells}
  \sum_\tau I(P_\tau, C^*_\tau) =  
  O\left( D^3 \cdot \left( m^{3/2} / D^{9/2} + n / D^2 \right) \right) = 
  O\left( m^{3/2} / D^{3/2} + nD \right) . 
\end{equation}

We choose $D = am^{3/5} / n^{2/5}$, for a sufficiently small constant $a$. As 
before, we require that $1 \le D \le a' \min \{m^{1/3}, n^{1/2}\}$, for another 
sufficiently small constant $a' > 0$, or that
$b_1 n^{2/3} \le m \le b_2 n^{3/2}$, for suitable constants $b_1, b_2$. The 
analysis now proceeds exactly as in Section~\ref{ch:unit_circles}, and implies 
that the overall incidence bound, over all possible ranges of $m$, within the 
cells, is $O(m^{3/5}n^{3/5} + m + n)$.

Consider next incidences involving points that lie on $Z(f)$. A lifted circle
$c^*$ that is not fully contained in $Z(f)$ crosses it in at most $O(D)$ points 
(since it is a constant-degree algebraic curve), for an overall $O(nD)$ bound, 
which is asymptotically subsumed by the bound (\ref{eq:tangent_in_cells}) within 
the cells. It therefore remains to bound the number of incidences between the 
points of $P$ on $Z(f)$ and the lifted circles that are fully contained in
$Z(f)$. 

Handling points and curves on $Z(f)$ is done as in the proof of Theorem 1.4~in
\cite{SharirSolomon} and in Section~\ref{ch:unit_circles}. We recall that it 
considers each irreducible component of $Z(f)$ separately, and distinguishes 
between the case where the component is \emph{infinitely ruled} by lifted 
circles, and the case where it is not. Fortunately for us, as in the previous 
section, the first case cannot arise, as we show next. 
\begin{restatable}{lemma}{lemtangentnoinfruled}\label{lem:tangent_no_inf_ruled}
  No algebraic surface is infinitely ruled by lifted circles.
\end{restatable}

\noindent{\bf Proof.}
Assume to the contrary that there exists an algebraic surface $V$ that is 
infinitely ruled by curves from $\C$; assume also, without loss of generality, 
that $V$ is irreducible (again, see a previous footnote for the justification of
this assumption). Write $V$ as the zero set $Z(f)$ of an irreducible 
trivariate real polynomial $f$.

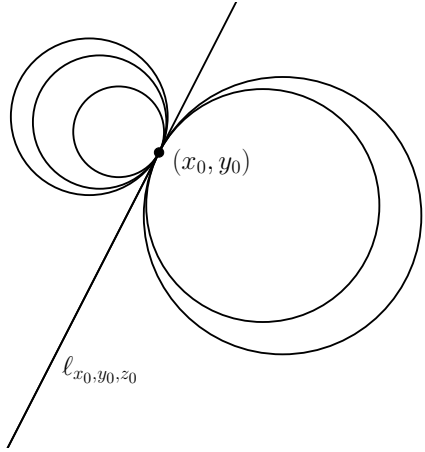
\begin{figure}[htb]
  \begin{center}
    \resizebox{6cm}{!}{\input{figures/infruled.tikz}}
    \caption{ \small \sf The infinite family $\C_V(x_0,y_0,z_0)$ consists of 
    lifted circles that are incident to the point $(x_0,y_0,z_0)$ and are 
    contained in $V$. The corresponding set $C_V(x_0,y_0,z_0)$ of circles in the 
    plane are all incident to the directed point $((x_0, y_0), z_0)$, and a 
    sample of these circles are depicted in the figure. The line
    $\ell_{x_0,y_0,z_0}$ that passes through $(x_0,y_0)$ and has slope $z_0$ is 
    tangent to all these circles.} 
    \label{fig:infruled}
  \end{center}
\end{figure}

Fix a point $(x_0, y_0, z_0) \in V$, and let $\C_V(x_0, y_0, z_0) \subseteq \C$ 
denote the set of all lifted circles in $\C$ that are incident to
$(x_0, y_0, z_0)$ and are contained in $V$. Since we assume that $V$ is 
infinitely ruled by $\C$, we have that $\C_V(x_0, y_0, z_0)$ consists of 
infinitely many curves. By construction, back in the plane, all the infinitely 
many circles in $C_V(x_0, y_0, z_0) := \set{c}{c^* \in \C_V(x_0, y_0, z_0)}$ are
incident to the directed point $((x_0, y_0), z_0)$. Denote by
$\ell_{x_0, y_0, z_0}$ the line that passes through $(x_0, y_0)$ and has slope
$z_0$. See Figure~\ref{fig:infruled} for an illustration.

We claim that $C_V(x_0, y_0, z_0)$ contains all the circles that are incident to
$((x_0, y_0), z_0)$. Indeed, let $C'_{x_0, y_0, z_0}$ denote the family of all 
circles that are incident to $((x_0, y_0), z_0)$, and let
$V' = \bigcup \set{c^*}{c \in C'_{x_0, y_0, z_0}}$. We claim that
$V'$ is an irreducible algebraic surface of degree $3$.

Indeed, a circle $c \in C'_{x_0, y_0, z_0}$ with radius $c_R$ and center
$(c_x, c_y)$ has to satisfy the following two polynomial equations:
\[(x_0 - c_x)^2 + (y_0 - c_y)^2 = c_R^2 ,\quad\text{and}\quad 
z_0 (y_0 - c_y) = c_x - x_0 .\]
According to (\ref{eq:lifted_circle}), we can write $c^*$ as the zero set of
\[(x - c_x)^2 + (y - c_y)^2 = c_R^2 ,\quad\text{and}\quad 
z (y - c_y) = c_x - x .\]
Combining the last four equalities, and eliminating $c_x, c_y, c_R$, we arrive 
at the following polynomial of degree $3$:
\[
f(x, y, z) = 
(z + z_0)((y - y_0)^2 - (x - x_0)^2) - 2 (x - x_0) (y - y_0) (z z_0 - 1) ,
\]
and $V'$ is the zero set of $f(x, y, z)$. It is easy to verify that $f(x, y, z)$
is irreducible (e.g., by considering the monomials that are divisible by $z$).

If $V \cap V'$ is two-dimensional then, since $V'$ and $V$ are irreducible,
we must have $V \cap V' = V' = V$. This implies that
$C'_{x_0, y_0, z_0} = C_V(x_0, y_0, z_0)$, as asserted.

Otherwise, $V \cap V'$ is an (at most) one-dimensional curve of degree at most
$deg(V) \cdot deg(V') = O(deg(V))$. Hence, it can contain at most $O(deg(V))$ 
curves of $C_V(x_0, y_0, z_0)$, a contradiction to the assumed infinite 
ruledness of $V$ (at $(x_0, y_0, z_0)$). This establishes the claim.

The argument just given also shows that $V$ is equal to the union of the lifted
images of all the circles that are incident to $((x_0,y_0),z_0)$, and this
property holds for every $(x_0,y_0,z_0)\in V$. Moreover, for any such
$(x_0,y_0,z_0)$, the $xy$-projection of $V$ is the entire plane, 
except (if we only consider the affine part of $V$)
for the line $\ell_{x_0,y_0,z_0}$. Let $(x_1,y_1,z_1)$ be another point in $V$, 
such that $(x_1,y_1)\notin \ell_{x_0,y_0,z_0}$ and $z_0\ne z_1$. Then $(x_1,y_1,z_1)$ 
lies on some lifted circle $C^*_{01}$ of $\C_V(x_0,y_0,z_0)$. That is, there exists
a circle $C_{01}$ that is incident to both $(x_0,y_0,z_0)$ and $(x_1,y_1,z_1)$.
Now take a third point $(x_2,y_2,z_2)\in V$ so that $(x_2,y_2)$ does not lie 
on either of the lines $\ell_{x_0,y_0,z_0}$ and $\ell_{x_1,y_1,z_1}$,
and also does not lie on $C_{01}$. We can therefore find two additional circles
$C_{02}$ and $C_{12}$ such that $C_{02}$ (resp., $C_{12}$) is incident to both
$(x_0,y_0,z_0)$ and $(x_2,y_2,z_2)$ (resp., $(x_1,y_1,z_1)$ and $(x_2,y_2,z_2)$),
and all three circles are distinct.

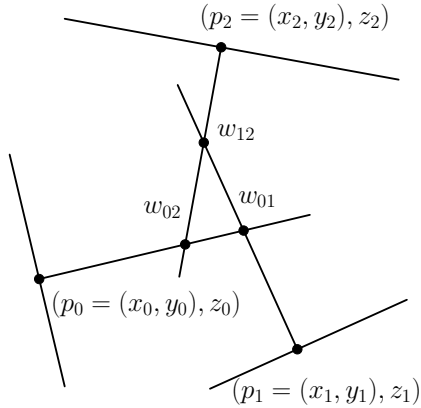
\begin{figure}[htb]
  \begin{center}
    \resizebox{6cm}{!}{\input{figures/fig012.tikz}}
    \caption{ \small \sf An impossible layout of three directed points $(x_0,y_0,z_0)$,
    $(x_1,y_1,z_1)$, and $(x_2,y_2,z_2)$ on $V$.}
    \label{fig012}
  \end{center}
\end{figure}

This however is impossible. To see this, denote by $w_{01}$ (resp., $w_{02}$, $w_{12}$)
the center of $C_{01}$ (resp., $C_{02}$, $C_{12}$). By construction, these points
are distinct and not collinear. Denote the point $(x_0,y_0)$ (resp., $(x_1,y_1)$, $(x_2,y_2)$)
as $p_0$ (resp., $p_1$, $p_2$). We then have (see Figure~\ref{fig012})
\begin{align} \label{eq:012}
p_0 w_{01} & = p_1 w_{01} \nonumber \\
p_1 w_{12} & = p_2 w_{12} \\
p_2 w_{02} & = p_0 w_{02} \nonumber .
\end{align}
But we have
\begin{align*}
p_1 w_{01} & = p_1 w_{12} + w_{12} w_{01} \\
p_2 w_{12} & = p_2 w_{02} + w_{02} w_{12} \\
p_0 w_{02} & = p_0 w_{01} + w_{01} w_{02} ,
\end{align*}
where the term $w_{12} w_{01}$ is signed---it is positive (resp., negative) if the 
vectors $\overrightarrow{p_1 w_{12}}$ and $\overrightarrow{w_{12} w_{01}}$ have
the same (resp., opposite) directions, and similarly for the terms
$w_{02} w_{12}$, $w_{01} w_{02}$. 

Substituting these equalities into (\ref{eq:012}) and adding up the three equations, we get
$$
w_{12} w_{01} + w_{02} w_{12} + w_{01} w_{02} = 0 ,
$$
which is impossible, no matter what signs these lengths have, since the triangle $w_{01} w_{02} w_{12}$ is non-degenerate.

We therefore reach a contradiction, concluding that no algebraic surface $V$ 
in $\reals^3$ can be infinitely ruled by curves of $\C$.
$\Box$\medskip

We can now continue exactly as in Section~\ref{ch:unit_circles}. Briefly, we 
consider each irreducible component of $Z(f)$ separately, enumerating them as
$Z(f_1), \ldots, Z(f_k)$, for suitable irreducible polynomials
$f_1, \ldots, f_k$, of respective degrees $D_1, \ldots, D_k$, where
$\sum_{i = 1}^k D_i \le D$. By what we have just shown, none of these components 
is infinitely ruled by lifted circles. Let $C_i^*$ denote the set of all lifted 
circles of $C^*$ that are fully contained in $Z(f_i)$. By 
Theorem~\ref{thm:inf_ruled_exceptional}, there exist absolute constants $c$,
$t$, such that there are at most $c D_i^2$ ``exceptional'' lifted circles in
$C_i^*$, namely, lifted circles that contain at least $c D_i$ $t$-rich points of
$P \cap Z(f_i)$. Denote the number of $t$-rich (resp., $t$-poor) points as $m_{rich}$ 
(resp., $m_{poor}$), so $m_{rich} + m_{poor} = m^* = |P\cap Z(f)|$. 
By choosing $a$ (in the definition of $D$) sufficiently small, we can ensure that
$\sum_i D_i^2 \le (\sum_i D_i)^2 \le D^2 \le n/(2c)$. 

As in Section~\ref{ch:unit_circles}, assign each point $p \in P\cap Z(f)$ 
(resp., lifted circle $c \in C^*$ contained in $Z(f)$) to the first irreducible 
factor $f_i$, such that $Z(f_i)$ contains $p$ (resp., fully contains $c$). 
The number of `cross-incidences', between points and lifted circles assigned to different 
components, is, as before, $O(nD)$, which satisfies our asymptotic bound. 

The number of `within-components' incidences on 
the non-exceptional lifted circles, summed over all components $Z(f_i)$, is
$O(m_{poor} + nD)$. Indeed, each non-exceptional lifted circle contains at most
$c D_i$ $t$-rich points, for a total of $O(n D_i)$ incidences, and the sum of
these bounds is $O(n D)$. Any $t$-poor point lies on at most $t$ lifted circles 
of $C_i^*$, for a total of $t m_{poor} = O(m_{poor})$ incidences (over all sets
$C_i^*$).

For the exceptional lifted circles, we simply apply induction, as their overall 
number, which is at most $c \sum_i D_i^2 \le c D^2 \le n/2$.
Note that in this inductive step we only need to consider the $t$-rich points, 
as the $t$-poor points have already been taken care of.
By the induction hypothesis, the corresponding incidence bound between the 
points and circles that were assigned to the same
$f_i$ is at most
\[
A \left( m_i^{3/5} (c D_i^2)^{3/5} + m_i + c D_i^2 \right) ,
\]
where $m_i$ is the number of $t$-rich points assigned to $f_i$. We now sum over
$i$. The second term is bounded by $m_{rich}$, since $\sum_i m_i = m_{rich}$. 
The third term is bounded by $n / 2$, since 
$c \sum_i D_i^2 \le n / 2$. As for the first term, we use Hölder's inequality:
\small
\begin{equation}
  \sum_i m_i^{3/5}(c D_i^2)^{3/5} = c^{3/5} \sum_i m_i^{3/5}D_i^{6/5} \le
  c^{3/5} \left( \sum_i m_i \right)^{3/5} \left( \sum_i D_i^3 \right)^{2/5} =
  c^{3/5} m^{3/5} \left( \sum_i D_i^3 \right)^{2/5} .
\end{equation}
\normalsize
Finally, using the fact that $\sum_i D_i^3 \le D^3$, we get the overall bound:
\begin{equation} \label{eq:tangent_on_zero}
  A \left( c^{3/5}m^{3/5}D^{6/5} + m_{rich} + n / 2 \right) \le
  A \left( \frac{m^{3/5}n^{3/5}}{2^{3/5}} + m_{rich} + n / 2 \right) ,
\end{equation}
since $c^{3/5}D^{3/5} \le (n / 2)^{3/5}$, as follows from the preceding 
inequalities.

We now add to this bound the bound for incidences within the cells, as well as 
the various other bounds involving points on $Z(f)$. Together we can upper bound 
all these bounds by
\[
B \left( m^{3/5}n^{3/5} + n + m_0 + m_{poor} \right) ,
\]
for a suitable absolute constant $B$. By choosing $A$ sufficiently large, the 
sum of this bound and the bound in (\ref{eq:tangent_on_zero}) is at most
\begin{equation} \label{eq:tangent_main}
  I(P,C) \le A \left( m^{3/5}n^{3/5} + m + n \right) .
\end{equation}
This establishes the induction step, and thereby completes the proof of 
Theorem~\ref{thm:tangent_main}, modulo the still missing proof of 
Lemma~\ref{lem:tangent_boot}, presented next.
$\Box$\medskip

\subsection{Proof of Lemma~\ref{lem:tangent_boot}} \label{sec:tangent_boot_proof}

We dualize the setup, exploiting the underlying geometry in the plane, by mapping 
each circle $c \in C$, with center $(\xi, \eta)$ and radius $r$, to the point 
$q_c = (\xi, \eta, \zeta)$, where $\zeta = r^2 - \xi^2 - \eta^2$, and by mapping 
each directed point $(p, u)$ to the locus $h_{p, u}$ of all dual points that 
represent circles that are incident to $(p, u)$. We claim that $h_{p, u}$ is a 
line. Indeed, for a circle $c$, with center $(\xi, \eta)$ and radius $r$, we 
have, by definition, that $c$ is incident to $(p, u)$ if and only if
\begin{align*}
& (\xi - p_1)^2 + (\eta - p_2)^2 = r^2, \quad \text{and} \\
& \left( \xi - p_1, \; \eta - p_2 \right) \; \text{is orthogonal to $\vec{u} = (1, u)$} ,
\end{align*}
where we write $p = (p_1, p_2)$. In other words, $h_{p, u}$ is the locus of all
points $(\xi, \eta, \zeta)$ that satisfy
\begin{align*}
\zeta & = - 2 p_1 \xi - 2 p_2 \eta + (p_1^2 + p_2^2) \\
& (\xi - p_1) + u (\eta - p_2) = 0 ;
\end{align*}
as this is an intersection of two (distinct, non-parallel) planes, the claim 
follows. Note that the second equation, within the $\xi\eta$-plane, is the equation 
of the line $\ell_{p, u}$ that passes through $p$ in direction orthogonal to $u$ 
(similar to the way it was defined in the proof of Lemma~\ref{lem:tangent_no_inf_ruled}).
That is, $\ell_{p, u}$ is the $\xi\eta$-projection of $h_{p, u}$. 
(A word of caution about the notation: The lines $h_{p, u}$ are all distinct, as is easily 
verified, but a line $\ell_{p, u}$ is shared by all directed points $(p, u)$ 
where $p$ lies on the line and $u$ is the slope of the line.) That is, all these
lines $h_{p, u}$ project to $\ell_{p, u}$.

In other words, we have reduced the problem to that of incidences between $n$ points
(those dual to the circles of $C$) and $m$ lines (the lines $h_{p, u}$, for
$(p, u) \in P$) in three dimensions. We can apply the result of \citet{GuthKatz} 
(see Theorem~\ref{thm:guth_katz}) for estimating the number of these incidences, and obtain
\begin{equation} \label{eq:tangent_guth_katz_bound}
  I(P, C) = I(C^*, H) = O\left( n^{1/2}m^{3/4} + n^{2/3}m^{1/3}q^{1/3} + n + m \right) , 
\end{equation}
where $H$ is the set of the dual lines $h_{p,u}$, and $q$ is the maximum number of lines
of $H$ that can lie in a common plane. This is a notable difference with the 
analysis in Section~\ref{ch:unit_circles}: There we showed that no surface is infinitely 
ruled by the dual curves, whereas here every plane is such a surface. Handling incidences on 
planes requires extra work, presented in the next subsection. 

\subsection{Coplanar lines} \label{subsec:tangent-coplanar_lines}

The gist of the analysis in this subsection is to control the value of $q$. For this, we\
distinguish between planes that contain at most $q$ lines of $H$, for a suitable threshold
value $q$, and those that contain more than $q$ lines. We handle the latter type of planes 
using a different technique that strongly exploits the geometry of the problem, and are then
left with a subproblem in which (\ref{eq:tangent_guth_katz_bound}) can be used.

Recall that if $c$ has center $z$ and radius $r$ then the \emph{power} of a point $w$ with
respect to $c$ is ${| wz |}^2 - r^2$. As is well known, and easy to see, the duality 
transform that we have used to map circles in $\reals^2$ to points in $\reals^3$ has the 
property that for each non-vertical plane $\pi$ in $\reals^3$ there exist a point $w$ in
$\reals^2$ and a power $\rho$, such that the point dual to a circle $c$ lies on $\pi$ if and 
only if $w$ has power $\rho$ with respect to $c$. Indeed, represent $\pi$ as the zero set of
$ax + by + z + d = 0$ (this is possible, since $\pi$ is non-vertical). Then choose
$w = (a/2, b/2)$ and $\rho = d + a^2/4 + b^2/4$. Now $w$ has power $\rho$ with respect to a 
circle $c$ with center $(u, v)$ and radius $r$ if and only if
$(u - a/2)^2 + (v - b/2)^2 - r^2 = \rho$, which is equivalent to the point
$(u, v, r^2 - u^2 - v^2)$ being on $\pi$, as asserted.

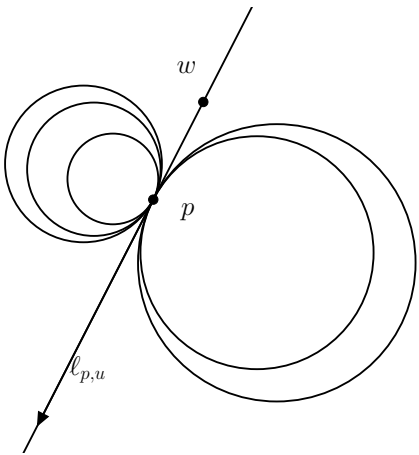
\begin{figure}[htb]
  \begin{center}
    \resizebox{6cm}{!}{\input{figures/power.tikz}}
    \caption{ \small \sf The locus of points $w$ that have a fixed power with respect to all
    the circles incident to a directed point $(p, u)$ is exactly the common tangent line
    $\ell_{p,u}$. Clearly, any point on $\ell_{p,u}$ has a fixed power (equal to the square 
    of its distance from $p$) with respect to all the circles incident to the directed point
    $(p, u)$; and any point not on $\ell_{p,u}$ will be inside some of these circles and 
    outside others (hence, having different powers with respect to them).} 
    \label{fig:power}
  \end{center}
\end{figure}

\begin{figure}[htb]
  \begin{center}
    \resizebox{6cm}{!}{\input{figures/power2.tikz}}
    \caption{ \small \sf A circle $\gamma = \gamma(w, \sqrt{\rho})$ and a point $p$ on
    $\gamma$ with direction $u$ to the center $w$ of $\gamma$. It is clear that any circle 
    incident to $(p,u)$ (such as the dashed circles) has power $\rho$ with respect to $w$.} 
    \label{fig:power2}
  \end{center}
\end{figure}
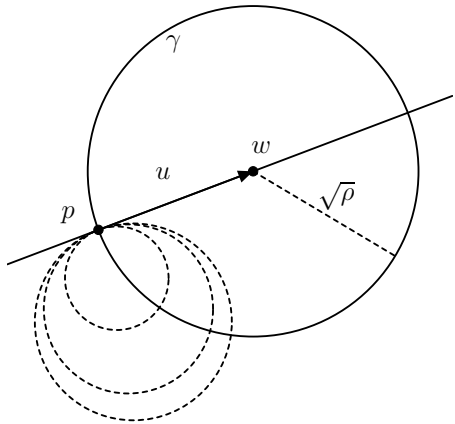

Let $\pi$ be any fixed non-vertical plane in $\reals^3$, and let $w$ and $\rho$ be the 
corresponding point and power (in $\reals^2$). It follows that a line $h_{p, u}$ is fully 
contained in $\pi$ if and only if all circles that are incident to $(p, u)$ have the fixed 
power $\rho$ with respect to the fixed point $w$. This is possible if and only if $w$ lies 
on the common tangent line to all the circles whose lifted images lie in $h_{p, u}$, that 
is, the line $\ell_{p, u}$ itself (see Figure~\ref{fig:power} for an illustration). The power
$\rho$ in this case is $|pw|^2 > 0$. We then denote $\pi$ as $\pi(w, \sqrt{\rho})$. In other 
words, all the lines $h_{p, u}$ that lie in a common non-vertical plane
$\pi(w, \sqrt{\rho})$, with respective parameters $w$ and $\rho > 0$, are such that (a) $p$ 
lies on the fixed circle $\gamma(w, \sqrt{\rho})$, with center $w$ and radius $\sqrt{\rho}$, 
and (b) $u$ is the direction of the line connecting $p$ and $w$ (see Figure~\ref{fig:power2}
for an illustration).

Let $P^+ = \{(p, u)^+ = (p, -1/u) \mid (p, u) \in P \}$, and let $\W$ denote the set of all 
possible ``power circles'' $\gamma(w, \sqrt{\rho})$ as just defined. By construction,
$h_{p, u}$ lies in a plane $\pi(w, \sqrt{\rho})$ if and only if $(p, u)^+$ is incident to 
the corresponding circle $\gamma(w, \sqrt{\rho})$.
A plane $\pi = \pi(\omega, \sqrt{\rho})$ contains $q$ lines $h_{p,u}$ if and only if
$\gamma(\omega, \sqrt{\rho})$ is incident to $q$ points of $P^+$. 

We fix a threshold value $q$, to be determined shortly, and partition $\W$ into two subsets
$\W^+$, $\W^-$, where $\W^+$ (resp., $\W^-$) consists of those circles in $\W$ that are 
incident to more than (resp., at most) $q$ directed points of $P^+$. We refer to circles in
$\W^+$ (resp., in $\W^-$) as being \emph{$q$-rich} (resp., \emph{$q$-poor}). The same 
notation carries over to the corresponding power planes in $3$-space.

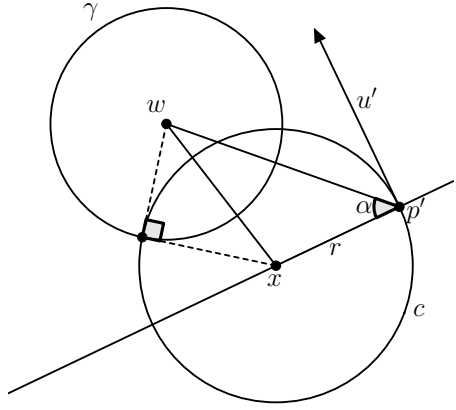
\begin{figure}[htb]
  \begin{center}
    \resizebox{6cm}{!}{\input{figures/unique1.tikz}}
    \caption{ \small \sf We have a circle $\gamma = \gamma(w, \sqrt{\rho})$ and 
    a directed point $(p', u')$ not incident to $\gamma$. A circle $c$ with 
    center $x$ and radius $r$ that is tangent to $\ell_{p', u'}$ at $p'$ and 
    orthogonal to $\gamma$, must have $x$ lie on the line through $p'$ that is 
    orthogonal to $u'$, and in addition must satisfy $r^2 + \rho = |wx|^2$. By 
    the Law of Cosines, we have
    $|wx|^2 = r^2 + |wp'|^2 - 2r|wp'|\cos\alpha$, where $\alpha = \angle wp'x$. 
    Combining these two 
    equalities, we get $2r|wp'|\cos\alpha = |wp'|^2 - \rho$. Since
    $\rho$, $|wp'|$, and $\alpha$ are fixed (depending only on $\gamma$,
    $p'$, and $u'$), it follows that $r$ is fixed too, unless
    $\cos\alpha = 0$, which cannot be the case because then $(p', u')^+$ 
    would have to be incident to $\gamma$, as is easily checked. We conclude 
    that there are at most two such circles $c$ (one on each side of $p'$).} 
    \label{fig:unique1}
  \end{center}
\end{figure}

We first get rid of the directed points $(p, u)$ such that $h_{p,u}$ lies in some $q$-rich 
plane; that is, their orthogonal points $(p, u)^+$ are incident to a $q$-rich 
circle in $\W^+$. Let $(p, u)$ be such a ``rich'' point (in the sense just 
defined). We pick a $q$-rich plane $\pi(w, \sqrt{\rho})$ that contains $(p, u)$; 
if there are several such planes, we pick one arbitrarily. Then, for each circle
$c \in C$ that is incident to $(p, u)$, charge each incidence of
$c$ with some other point $(p', u')$ for which $(p', u')^+$ is not incident to
$\gamma(w, \sqrt{\rho})$, assuming there are such points, to the pair
$\left( (p',u'),\; \pi(w, \sqrt{\rho}) \right)$. The charging is unique, up to
multiplicity $2$. That is, $(p', u')$ and $\pi(w, \sqrt{\rho})$ determine 
at most two circles $c$ that satisfy these properties, and thus determine 
up to two incidences. Indeed, there can be at most two circles $c$ that 
pass through $p'$, are tangent to $\ell_{p', u'}$, and are orthogonal to 
the circle $\gamma(w, \sqrt{\rho})$, provided that $(p', u')^+$ is not incident to 
$\gamma(w, \sqrt{\rho})$; see Figure~\ref{fig:unique1} for an illustration and a proof.) 
If this charging fails, $c$ can be incident to at most two points of $P$, namely $(p, u)$ 
and the other point $(p', u')$, where $p'$ is the unique second point of intersection of
$c$ and $\gamma(w, \sqrt{\rho})$ and $u'$ is the slope of the corresponding
tangent. In such a case, the relevant circles $c$ have at most $2n$ incidences in total.
In other words, the number of incidences involving circles that are incident to at 
least one rich point is $O(m|\W^+| + n)$.

It remains to estimate $\W^+$. We use the following simple approach. Cut $3$-space by
some generic plane $\pi_0$. Each plane in the set $\Pi^+$ of $q$-rich planes (namely, 
the planes corresponding to the $q$-rich circles in $\W^+$) appears in the 
cross-section as a line, and each line $h_{p, u}$ appears as a point. With a suitable 
generic choice of $\pi_0$, the cross-sectional lines are all distinct, the 
cross-sectional points are all distinct, and a line is incident to a point in the 
cross section if and only if the corresponding plane $\pi(w, \sqrt{\rho})$ contains 
the corresponding line $h_{p, u}$.

Letting $H$ denote, as above, the set of lines $h_{p, u}$, for $(p, u) \in P$, 
and using Theorem~\ref{thm:szemeredi_trotter}, we have
\[
q|\W^+| \le I(H, \Pi^+) = O\left( m^{2/3}|\W^+|^{2/3} + m + |\W^+| \right) ,
\]
or
\[
|\W^+| = O\left( \frac{m^2}{q^3} + \frac{m}{q} \right) .
\]
To recap, we have shown that the number of incidences in $I(P, C)$ involving circles 
that are incident to at least one rich point is 
\[
O(m|\W^+| + n) = O\left( \frac{m^3}{q^3} + \frac{m^2}{q} + n \right) .
\]
To complete this part of the analysis, we also have to consider vertical planes. 
Clearly, no line $h_{p, u}$ can be contained in more than one vertical plane (for 
otherwise $h_{p, u}$ would have to be vertical, which is impossible). Hence the 
number of $q$-rich vertical planes is at most $m / q$.

\begin{figure}[htb]
  \begin{subfigure}{.5\textwidth}
    \centering
    \resizebox{6cm}{!}{\input{figures/unique21.tikz}}
    \caption{\sf\small }
  \end{subfigure}%
  \begin{subfigure}{.5\textwidth}
    \centering
    \resizebox{6cm}{!}{\input{figures/unique22.tikz}}
    \caption{\sf\small }
  \end{subfigure}
  \begin{center}
    \caption{ \small \sf (a) Given a directed point $(p', u')$ and a line
    $\ell_{p, u}$ distinct from $\ell_{p',u'}$, there are at most two circles $c$ 
    incident to $(p', u')$ and tangent to $\ell_{p,u}$. (b) If $\ell_{p,u}$ coincides 
    with $\ell_{p',u'}$ but $p \neq p'$, no circle can be tangent to both directed
    points.} 
    \label{fig:unique2}
  \end{center}
\end{figure}
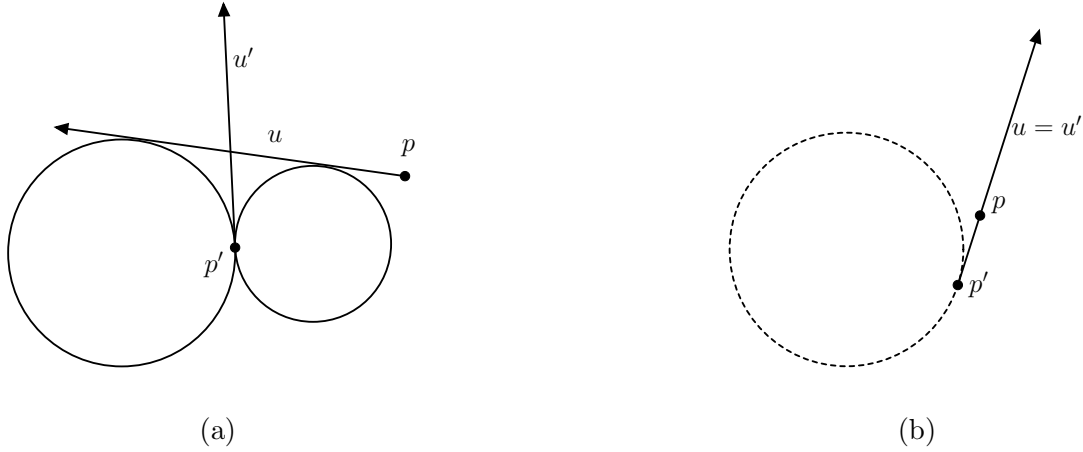

In this case we say that a point $(p, u)$ is \emph{rich} if the vertical plane containing
$h_{p, u}$ is $q$-rich. We charge an incidence of a point $(p', u')$ with any
circle $c \in C$ that is also incident to at least one rich point $(p, u)$, to 
the vertical plane containing $h_{p, u}$, or, alternatively, to the line
$\ell_{p, u}$ in the $xy$-plane (which is the cross-section of that vertical 
plane with the $xy$-plane). Here too the charging is almost unique: The point
$(p', u')$ and the line $\ell_{p, u}$ determine at most two circles with these 
properties---there can be at most two circles that are incident to $(p', u')$ 
and are tangent to $\ell_{p, u}$, and this holds since the lines $\ell_{p, u}$ 
and $\ell_{p', u'}$ do not coincide; see Figure~\ref{fig:unique2}. There is at 
most one uncharged incidence on each circle, for an additional count of at most
$n$.

Hence, the analysis proceeds more or less exactly as in the preceding case, with 
the exception that the term $m^3 / q^3$ does not arise, and we only get the 
terms $O(m^2/q + n)$. 

We discard all rich points (of either of the two types) and all their incident 
circles; the deleted incidences have all been accounted for.

The remaining incidences only involve the surviving circles (we continue to 
denote their set as $C$ and its size as $n$) and the $q$-poor points; we denote 
their set as $P_{poor}$ and still denote its size as $m$. By construction, we 
now have that, in the dual 3-space, no plane contains more than $q$ lines
$h_{p, u}$.

Applying Guth and Katz's bound~\cite{GuthKatz} in the reduced scenario, the 
number of surviving incidences is thus
\[
O\left( n^{1/2}m^{3/4} + n^{2/3}m^{1/3}q^{1/3} + n + m \right) ,
\]
for a total of
\begin{equation} \label{weakincs}
O\left( n^{1/2}m^{3/4} + n^{2/3}m^{1/3}q^{1/3} + n + \frac{m^3}{q^3} + \frac{m^2}{q} 
\right) 
\end{equation}
incidences (the term $m$ is always subsumed by the last term, since $q \le m$).

We now choose $q$. We have two options for doing this. We first balance the second 
and fourth terms in (\ref{weakincs}), and choose $q = m^{4/5} / n^{1/5}$ accordingly.
For this to be the right choice, we require that (a) $1 \le q \le m$, and (b) the 
fifth term is dominated by, say, the fourth, that is, $q \le m^{1/2}$. The resulting 
inequalities
$1 \le q \le m^{1/2}$ hold when $n^{1/4} \le m \le n^{2/3}$. We can always assume that
$m \ge n^{1/4}$, for otherwise even the naive bound $O(m^2 + n)$ yields only $O(n)$ 
incidences. That is, if $m \le n^{2/3}$ we obtain the bound
\[
O\left( n^{1/2}m^{3/4} + n^{3/5}m^{3/5} + n \right) =
O\left( n^{1/2}m^{3/4} + n \right) ,
\]
because the second term is dominated by the third term in this range.

The other option is to balance the second and fifth terms in (\ref{weakincs}), and 
thus choose $q = m^{5/4} / n^{1/2}$. Again, for this to be the right choice, we require 
that (a) $1 \le q \le m$, and (b) the fourth term is dominated by the fifth, that is,
$q \ge m^{1/2}$. The resulting inequalities $m^{1/2} \le q \le m$ hold when
$n^{2/3} \le m \le n^2$. Since we only consider directed points that are incident to
at least two circles of $C$, the results of Ellenberg et al.~\cite{ESZ} imply that
$m = O(n^{3/2}) = O(n^2)$ (see Theorem~\ref{thm:ellenberg}). That is, if
$m \ge n^{2/3}$ we again obtain the bound
\[
O\left( n^{1/2}m^{3/4} + n \right) ,
\]
as is easily checked. That is, the number of incidences is always
$O\left( n^{1/2}m^{3/4} + n \right)$. Note that this is upper bounded by
$O(n + m^{3/2})$. That is, we have managed to improve the naive bound $O(n + m^2)$
to $O(n + m^{3/2})$, thus completing the proof of Lemma~\ref{lem:tangent_boot}, and, 
consequently, also of Theorem~\ref{thm:tangent_main}.
$\Box$\medskip


\section{Generalizations} \label{ch:generalizations}

\subparagraph*{The setup.}

In both problems studied in Sections~\ref{ch:unit_circles} and
\ref{ch:point_circle}, we employ the same technique, with similar intermediate 
lemmas, and obtain the same incidence bounds --- see 
Theorems~\ref{thm:anchored_main} and \ref{thm:tangent_main}. In this section we
present a general approach, extending those in Sections~\ref{ch:unit_circles} 
and \ref{ch:point_circle}, for bounding the number of incidences between points 
and curves in three dimensions that have almost two degrees of freedom, and 
satisfy some additional assumptions.

There is another recent work in progress by Guth et al., which
shows that the number of incidences between $m$ points and $n$ lines in $\reals^3$
that are taken from a family of lines that have almost two degrees of freedom and are
$3$-parameterizable, is also $O(m^{3/5}n^{3/5} + n)$, under certain natural assumptions
on the lines. Although the analysis that leads to Guth et al.'s bound is similar 
to the ones given in Sections~\ref{ch:unit_circles} and \ref{ch:point_circle}, there is
one major difference, in that in their scenario there can exist surfaces (namely,
planes) that are infinitely ruled by such a family of lines. We will use their setup 
as an example of the generalized analysis derived in this section.

Spelled out in more detail, the common characteristic of the problems studied in the
two previous sections is that they essentially deal with incidences between a set $P$
of points and a set $C$ of curves in three dimensions, where the curves of $C$ are
taken from a family $\C$ of curves that (i) have almost two degrees of freedom, and
(ii) are 3-parameterizable (that is, each of them is defined in terms of three real
parameters; see below for a precise formulation). Additionally, we need to assume that
not all pairs of curves of $\C$ intersect, which typically is the case, and a few 
additional (natural) assumptions --- see below.

A natural target would thus be to show that, under assumptions of this kind,
\[I(P, C) = O(|P|^{3/5}|C|^{3/5} + |P| + |C|) .\]

Unfortunately, this does not hold in general. As a simple counter-example, let $\C$
be the family of all horizontal lines in $\reals^3$ (i.e., parallel to the $xy$-plane);
this is a special (and rather degenerate) case of the setup studied by Guth et al.: 
It is easy to verify that $\C$ has almost two degrees of freedom (the corresponding
$6$-variate polynomial $F(p, q)$ is simply $p_z - q_z$),  that the lines of $\C$ are
$3$-parameterizable, and that not every pair of horizontal lines intersect. However, 
we can still place $m$ points and $n$ lines (from $\C$) in a single horizontal plane, 
so that they form $\Theta(m^{2/3}n^{2/3} + m + n)$ incidences, which is larger that the
desired bound.

Comparing this example with the analysis in the preceding sections, we see that, as
already noted, a major difference between the setups studied there and the one in 
this example is that in the previous cases there does not exist a surface that is 
infinitely ruled by curves from $\C$, whereas here every horizontal plane is such a 
surface.

In the analysis presented in this section we will need to address situations in which
there exist surfaces that are infinitely ruled by the curves of $\C$. As a matter of 
fact, as in previous sections, our analysis also involves a stage where it studies the
problem in a dual setting, and the existence or non-existence of infinitely ruled surfaces 
is an issue that has to be faced in this setting too, as it was in the preceding sections.

In more detail, since the curves in $\C$ are $3$-parameterizable, we can transform
the problem into a dual three-dimensional setting, where the curves become points 
and, as is easily checked, under natural assumptions, the points become curves, and 
we will make use of both the primal and the dual setups.

The reason why surfaces that are infinitely ruled by curves of $\C$ (and analogous
surfaces in the dual context) are of significance, is that we will need to handle 
incidences involving points and curves that lie on infinitely ruled components of
the partitioning polynomials, both in the primal and the dual settings, if such 
components exist, and the number of such incidences can be larger than then desired bound,
as the example with horizontal lines indicates. Fortunately, these issues either did not
exist, or were relatively easy to handle, in the problems studied in 
Sections~\ref{ch:unit_circles} and \ref{ch:point_circle}. However, in more general
contexts we cannot rule out the existence of such surfaces, and a special treatment of this 
kind of surfaces will therefore be required.

\subparagraph*{The general setup.} 
In more rigorous terms, we say that a family of curves $\C$ is
\emph{$3$-parameterizable} if each of them is specified in terms of three real
parameters. Concretely, catering to the case where the curves of $\C$ are given in 
parametric form, there exist three constant-degree real $4$-variate polynomials
$H_1(t; a, b, c), H_2(t; a, b, c), H_3(t; a, b, c)$, so that for each curve
$\gamma \in \C$ there exists a triple $(a_\gamma, b_\gamma, c_\gamma)$ of parameters,
so that $\gamma$ is given in parametric form (in terms of $t$) by
\begin{align*}
& x = H_1(t; a_\gamma, b_\gamma, c_\gamma) \\
& y = H_2(t; a_\gamma, b_\gamma, c_\gamma) \\
& z = H_3(t; a_\gamma, b_\gamma, c_\gamma), \quad\text{for } t \in \reals.
\end{align*}
A similar definition can be given when each curve $\gamma$ is given as the common 
zero set of two polynomials (in $(x, y, z)$), but we stick, for simplicity, and 
with some small loss of generality (in assuming that the $H_i$'s are polynomials), 
to the parametric setup. We then write the curve $\gamma$ as
$\gamma_{a,b,c}$, for $a = a_\gamma, b = b_\gamma$, and $c = c_\gamma$.

Note that both of the families of curves that we discussed in 
Sections~\ref{ch:unit_circles} and \ref{ch:point_circle} are $3$-parameterizable 
(as is the family of all horizontal lines). 

Throughout this section, let $\C$ be a family of constant-degree irreducible 
algebraic curves in $\reals^3$, so that (i) the curves in $\C$ have almost two 
degrees of freedom, and (ii) they are $3$-parameterizable. Let $F$ be the polynomial 
such that $p, q$ admit a curve of $\C$ that passes through both of them if and only 
if $F(p, q) = 0$. Assume further that $F$ is irreducible, an assumption that can be 
made without loss of generality, as argued in Section~\ref{subsec:gen-irreducible} below.

We recall that the technique described in Sections~\ref{ch:unit_circles} and
\ref{ch:point_circle}, that we wish to generalize in this section, proceeds as 
follows. We first establish a bootstrapping bound, which is weaker than the 
bound we are after, but stronger than the naive bound for curves with two 
degrees of freedom; we do this in Section~\ref{subsec:gen-bootstrapping}. In 
establishing this bound, we pass to a dual $3$-space, where the points become 
curves and the curves become points, and argue that both properties, of having 
at most two degrees of freedom and being $3$-parameterizable, also hold for the 
family of the dual curves (under certain natural assumptions, detailed below). 
The bootstrapping bound is derived using the standard polynomial partitioning 
technique, in the dual setup, where a major step has to deal with incidences 
with dual curves that are contained in irreducible components of the zero set of 
the partitioning polynomial (in the dual), that are infinitely ruled by the 
family of dual curves.

We then apply, back in the primal space, a similar partitioning technique with a 
suitable polynomial $f$. Here too, a major step is the handling of incidences with
curves that are contained in irreducible components of the zero set of $f$ that are 
infinitely ruled by $\C$.

We were fortunate in both cases studied in Sections~\ref{ch:unit_circles} and
\ref{ch:point_circle}, in that we had a lemma --- 
Lemmas~\ref{lem:anchored_no_inf_ruled} and \ref{lem:tangent_no_inf_ruled}, respectively ---
that showed that there are no surfaces in the primal space that are infinitely ruled by
$\C$. Handling such surfaces in the dual space required some ad-hoc machinery that 
exploited the specific geometry of the problems.

The derivation of the bootstrapping bound is presented in
Section~\ref{subsec:gen-bootstrapping}. The full details of the application of the 
partitioning technique, and the resulting derivation of the upper bound on the 
number of incidences, are given in Section~\ref{subsec:gen-partitioning}. A 
specialization of the technique presented in this section to the case of (not
necessarily horizontal) lines in three dimensions that have almost two degrees 
of freedom and are $3$-parameterizable will be given in 
Section~\ref{subsec:gen-lines}.

\subsection{Duality and irreducibility} \label{subsec:gen-irreducible}

\subparagraph*{The primal setup.} 
Let $C$ be a set of $n$ irreducible curves from a family $\C$ that has almost two 
degrees of freedom and is $3$-parameterizable, and let $P$ be a set of $m$ points in
$\reals^3$. Let $F$ be the $6$-variate polynomial such that $p, q$ admit a curve of $\C$ that 
passes through both points if and only if $F(p, q) = 0$. Let $F = F_1 F_2 \cdots F_k$ 
be the decomposition of $F$ into its irreducible factors. 

Let $\gamma$ be a curve in $\C$, and fix a point $p_0 \in \gamma$. For each
$q \in \gamma$ (different from $p_0$) we have $F(p_0, q) = 0$, so there exists (at
least) one index $j = 1, \ldots, k$ such that $F_j(p_0, q) = 0$; write
$j_{p_0,q} = j$. Then there are infinitely many points $q \in \gamma$ that have the
same index $j_{p_0,q}$; call this index $j_{p_0}$. Then we must have
$F_{j_{p_0}}(p_0, q) \equiv 0$, over $q \in \gamma$ (because the intersection
$Z(F_{j_{p_0}}(p_0, q))\cap\gamma$ is either a curve or a finite set of points). 
It follows that there are 
infinitely many points $p_0 \in \gamma$ that share the same index $j_{p_0}$, call
it $j_\gamma$. The same argument as above implies that $F_{j_\gamma}(p,q) = 0$ for 
all $p, q \in \gamma$. Now split $\C$ into $k$ subfamilies $\C_1, \ldots, \C_k$, 
where $\C_j = \set{\gamma \in \C}{j_\gamma = j}$, and note that, by definition, 
each $\C_j$ has almost two degrees of freedom, and the polynomial $F_j$ associated 
with each $\C_j$ is irreducible. Clearly, if $C$ is a set of $n$ curves from $\C$ 
then we can write
$C = \bigcup_{j=1}^k C_j$, where $C_j = C \cap \C_j$, and we have 
\[
I(P, C) \le \sum_{j = 1}^{k}{I(P, C_j)}.
\]
That is, we argued that it suffices to restrict ourselves to the case where the 
polynomial $F$ that defines the almost two degrees of freedom property is irreducible.
(More precisely, since $k$ is constant, obtaining a bound for the irreducible case
yields the same bound for the general case, multiplied by $k = O(1)$.)

\subparagraph*{The dual setup.} 

We dualize the setup, exploiting the $3$-parametrization of $\C$. It is somewhat
simpler to present this duality under the assumption that each curve $\gamma \in \C$ 
is given as
\[\gamma = \set{p \in \reals^3}{f(p; u_\gamma, v_\gamma, w_\gamma) = 
g(p; u_\gamma, v_\gamma, w_\gamma) = 0},\] 
for $6$-variate distinct algebraic irreducible polynomials $f, g$,
which is general enough for our purposes.
We then dualize each curve $\gamma$ to the point
$\gamma^* = (u_\gamma, v_\gamma, w_\gamma)$ in $3$-space. We denote by $\C^*$ 
the family of all these dual points $\gamma^*$ (over all curves $\gamma \in \C$). We
also map each point $p \in P$ to the dual locus $p^*$ of all dual points $\gamma^*$ 
that represent curves $\gamma \in \C$ that are incident to $p$, that is,
\[
p^* = 
\set{(u_\gamma, v_\gamma, w_\gamma) \in \reals^3}{f(p; u_\gamma, v_\gamma, w_\gamma) = g(p; u_\gamma, v_\gamma, w_\gamma) = 0}.
\]
Clearly, $p^*$ is an intersection of the zero sets of the two polynomials
$f(p; u, v, w)$ and $g(p; u, v, w)$ in $\reals[u, v, w]$, and is therefore a 
one-dimensional curve, unless these polynomials have a common factor. 
Note that, even when $f$ and $g$ have no common factor as four-variate polynomials
(being distinct and irreducible), there might exist points $p$ for which
$f(p; u, v, w)$ and $g(p; u, v, w)$, as  polynomials in $u, v, w$ with $p$ fixed, 
have a common factor.

As an illustration of this difficulty, consider the family $\L$ of all the lines in
$\reals^3$ that meet the $x$-axis. This family has almost two degrees of freedom,
and is $3$-parameterizable. However, for any point $p$ on the $x$-axis, $p^*$ is
not a curve in the dual space, but a two-dimensional surface, as is easily checked 
($p^*$ is a curve for any other point $p$).

We will therefore assume that $P$ is a set of points in $\reals^3$ for which $p^*$ is a 
one-dimensional curve. One can show, as in von zur Gathen~\cite[Lemma 4.3]{Von},
that for most points $p$, except for those that lie in some lower-dimensional variety,
$p^*$ is indeed a curve. 
(Informally, irreducibility of a polynomial is some algebraic 
condition in its coefficients. Generically, polynomials are irreducible (when the
coefficients are generic), but when one specializes values for some
coefficients, making them algebraically dependent, the polynomial can become reducible.
Again, see \cite{Von} for details.)
We note that in the cases considered in Sections~\ref{ch:unit_circles} and
\ref{ch:point_circle}, $p^*$ is a curve for all points $p$ (in the case of anchored unit 
circles in $\reals^3$, this is true for all $p$ other that the origin, and, to stay in the 
real domain, $p$ has to lie in the ball $\norm{p} \le 2$).

Recall that we have assumed that not all pairs of curves of $\C$ intersect. We claim 
that, under these assumptions, the family $\P^*$ of such dual curves (i) indeed consists
of one-dimensional curves, (ii) has almost two degrees of freedom, and (iii) is 
$3$-parameterizable. Property (i) holds automatically, since we have restricted $P$
to consist only of such points. Property (iii) is trivial, since the three coordinates 
of $p$ serve as the three parameters that specify $p^*$. For property (ii), let
$\gamma_1 = \gamma_{a_1,b_1,c_1}$ and
$\gamma_2 = \gamma_{a_2,b_2,c_2}$ be two distinct curves in $\C$. The condition
that there exists a dual curve $p^*$ that passes through both $\gamma_1^*, \gamma_2^*$
is the dual version of the condition that $\gamma_1$ and $\gamma_2$ intersect
(at the primal point $p$). This primal property means that there exist
$t_1, t_2 \in \reals$ such that (now we switch back to the parametric representation,
to simplify the presentation)
\[
H_j(t_1; a_1, b_1, c_1) = H_j(t_2; a_2, b_2, c_2),
\]
for $j = 1, 2, 3$. By eliminating $t_1$ and $t_2$ from these three equations (see
\cite{CLO}), we get a polynomial, denoted as $F^*$, satisfying 
\[
F^*(a_1, b_1, c_1; a_2, b_2, c_2) = 0.
\]
As long as $F^*$ does not vanish identically, we conclude that the dual system 
does indeed have almost two degrees of freedom. The converse implication, that 
if $F^*=0$ then $\gamma_1$ and $\gamma_2$ intersect, also follows from the 
properties of resultants (see \cite{CLO}). If $F^*$ were identically zero, then 
every pair of curves of $\C$ would have to intersect, contrary to assumption.

We can therefore apply exactly the same reasoning as in the primal setup, and 
conclude that we may also assume that $F^*$ is irreducible. 

\subsection{Bootstrapping bound} \label{subsec:gen-bootstrapping}

As in the preceding sections, we derive an improved bootstrapping bound 
(over the naive bound $O(m^2+n)$), using the fact that a generic pair of 
points has no curve from $\C$ incident to both of them, since $\C$ has almost 
two degrees of freedom.  Concretely, the argument proceeds as follows. 

We dualize the setup, as described in Section~\ref{subsec:gen-irreducible}. Let $\P^*$ 
denote the family of all dual curves $p^*$ for points $p \in \reals^3$ for which the dual 
object is indeed a curve. As argued in Section~\ref{subsec:gen-irreducible}, the family
$\P^*$ has almost two degrees of freedom and is $3$-parameterizable.

Let $C^* \subset \C^*$ be the set of points $c^*$ dual to the curves $c \in C$, 
and let $P^* \subset \P^*$ be the set of curves $p^*$ dual to the points
$p \in P$ (by assumption they are indeed curves). We have thus reduced our 
problem to that of bounding the number of incidences between a set $C^*$ of $n$ 
points and a set $P^*$ of $m$ curves in $\reals^3$, taken from some family
$\P^*$ that has almost two degrees of freedom, and is (naturally)
$3$-parameterizable (by the coordinates of the corresponding primal points). 
We can bound the number of 
incidences using Theorem~\ref{thm:dof_3d}, exploiting the fact that the curves 
of $\P^*$ have two degrees of freedom. This yields the bound
\begin{equation} \label{eq:gen_thm14_bound}
  I(P, C) = I(C^*, P^*) =
   O\left( n^{1/2}m^{3/4} + n^{2/3}m^{1/3}\delta^{1/3} + n + m \right)
  , 
\end{equation}
where $\delta$ is the maximum number of curves from $P^*$ that lie on a common 
surface that is infinitely ruled by the family of dual curves $\P^*$. As this
can be upper bounded as $I(P, C) = O(m^{3/2} + n + n^{2/3}m^{1/3}\delta^{1/3})$,
we obtain:

\begin{lemma} \label{lem:gen_bootstrap}
  Let $C$ be a set of $n$ curves in $\reals^3$ that are taken from a $3$-parameterizable
  family $\C$ with almost two degrees of freedom, and let $P$ be a set of $m$ points in
  $\reals^3$. Let $\P^*$ be the family of curves in dual $3$-space (with respect to the 
  curves of $\C$) that are dual to the points of $\reals^3$ (for which the duals are 
  indeed curves), and assume that no surface that is infinitely ruled by curves of
  $\P^*$ contains more than $\delta$ curves dual to the points of $P$, and that not all 
  pairs of curves of $\C$ intersect. Then
  \[
  I(P, C) = O(m^{3/2} + n + n^{2/3}m^{1/3}\delta^{1/3}).
  \]
\end{lemma}

\subsection{Polynomial partitioning and the incidence bound} \label{subsec:gen-partitioning}

We now use the bootstrapping bound from Lemma~\ref{lem:gen_bootstrap} to prove a 
stronger bound, following the general approach in \cite{SharirSolomon}, and in the two
preceding sections. For convenience, we reproduce it here again, at the risk of some repetition.
Concretely, using induction on $n$, we will show that
\begin{equation} \label{eq:gen_induction}
  I(P, C) \le 
  A \left(m^{3/5}n^{3/5} + (m^{11/15}n^{2/5} + n^{8/9})\delta^{1/3} +
  m^{2/3}n^{1/3}\pi^{1/3} + m + n \right),
\end{equation}
for some suitable constant $A$, where $\delta$ (resp., $\pi$) is an upper bound on the
number of dual curves $p^* \in P^*$ (resp., primal curves $\gamma \in C$) that lie on a 
surface that is infinitely ruled by dual curves of $\P^*$ (resp., curves of $\C$).

The induction hypothesis clearly holds for $n \le n_0$, for some suitable threshold constant 
$n_0$, by making $A$ sufficiently large. For the induction step, we construct (again,
using Theorem~\ref{thm:polynomial_part})
a partitioning polynomial $f$ in $\reals^3$, of some specified (maximum) degree $D$, so that
each cell (connected component) of $\reals^3 \setminus Z(f)$ contains at most $O(m / D^3)$ 
points of $P$, and is crossed by at most $O(n / D^2)$ curves of $C$; the existence of such 
a polynomial follows from Theorem~\ref{thm:polynomial_part}.

For each (open) cell $\tau$ of the partition, let $P_\tau$ denote the set of points of 
$P$ inside $\tau$, and let $C_\tau$ denote the set of curves of $C$ that cross $\tau$. 
We have $m_\tau := |P_\tau| = O(m / D^3)$, and 
$n_\tau := |C_\tau| = O(n / D^2)$. We apply the bootstrapping bound of 
Lemma~\ref{lem:gen_bootstrap} within each cell $\tau$, to obtain
\small
\[
I(P_\tau,C_\tau) = 
O\left( \left(\frac{m}{D^3}\right)^{3/2} +
 \left(\frac{n}{D^2}\right)^{2/3}\left(\frac{m}{D^3}\right)^{1/3}\delta^{1/3} + 
 \frac{n}{D^2} \right) = 
O\left( \frac{m^{3/2}}{D^{9/2}} + \frac{n^{2/3}m^{1/3}\delta^{1/3}}{D^{7/3}} + 
 \frac{n}{D^2} \right) .
\]
\normalsize
Multiplying by the number of cells, we get that the number of incidences within the
cells is
\small
\begin{equation} \label{eq:gen_in_cells}
  \sum_\tau I(P_\tau,C_\tau) = 
  O\left( D^3\cdot \left( \frac{m^{3/2}}{D^{9/2}} + \frac{n^{2/3}m^{1/3}\delta^{1/3}}{D^{7/3}} + \frac{n}{D^2} \right) \right) = 
  O\left( \frac{m^{3/2}}{D^{3/2}} + n^{2/3}m^{1/3}\delta^{1/3}D^{2/3} + nD \right) . 
\end{equation}
\normalsize

As in the previous sections, we choose $D = am^{3/5}/n^{2/5}$, for a sufficiently small 
constant $a$. For this to make sense, we require that
$1 \le D \le a' \min\{m^{1/3}, n^{1/2}\}$, for another sufficiently small constant
$a' > 0$, which holds when $b_1 n^{2/3} \le m \le b_2 n^{3/2}$, for suitable constants
$b_1, b_2$ that depend on $a'$. 

If $m < b_1 n^{2/3}$, the bound in Lemma~\ref{lem:gen_bootstrap} yields (for the entire
sets $P$, $C$) the bound
\begin{equation} \label{eq:gen_small_m}
 O(m^{3/2} + n^{2/3}m^{1/3}\delta^{1/3} + n) = O(n^{8/9}\delta^{1/3} + n). 
\end{equation}

If $m > b_2 n^{3/2}$, we construct a partitioning polynomial $f$ of degree
$D = a'n^{1/2}$, for the same sufficiently small constant $a'$, so that each cell of
$\reals^3 \setminus Z(f)$ contains at most $O(m / D^3)$ points of $P$ and is 
crossed by at most $O(n / D^2) = O(1)$ curves of $C$ (we choose $n_0$ 
sufficiently large to ensure that $D > 1$ in this case). The number of 
incidences within each cell is then at most $O(m / D^{3})$, for a total of
$O(D^3) \cdot O(m / D^3) = O(m)$ incidences. More precisely, we write this bound
as $O(m_0)$, where $m_0$ is the number of points of $P$ within the cells. We 
also denote by $m^*$ the number of points of $P \cap Z(f)$, so $m_0 + m^* = m$.
Handling incidences on the zero set $Z(f)$ is done as in the case of 
a smaller $m$, as will be detailed shortly.

Assuming then that $m$ is in the middle range, and substituting
$D = am^{3/5} / n^{2/5}$, we get within the cells the bound
$I(P, C) = O(m^{3/5}n^{3/5} + m^{11/15}n^{2/5}\delta^{1/3})$. Hence, adding the bounds 
from the three subcases, we get, within the cells, the bound
\begin{equation} \label{eq:gen_in_cells_final}
 I(P, C) \le B\left(m^{3/5}n^{3/5} + (m^{11/15}n^{2/5} + n^{8/9})\delta^{1/3} + m_0 + n\right), 
\end{equation}
for some absolute constant $B$. We remark that the first $\delta$-dependent term 
dominates the second one if and only if
$m \ge n^{2/3}$.

Consider next incidences involving points that lie on $Z(f)$. A curve $c \in C$ 
that is not fully contained in $Z(f)$ crosses it in at most $O(D)$ points, 
since $c$ is a constant-degree algebraic curve. 
This yields a total of $O(nD)$ incidences, a bound that is asymptotically 
subsumed by the bound (\ref{eq:gen_in_cells_final}) for incidences within the 
cells. It therefore remains to bound the number of incidences between the points 
of $P$ on $Z(f)$ and the curves that are fully contained in $Z(f)$. Denote the 
subsets of these points and curves as $\widetilde{P}$ and $\widetilde{C}$, 
respectively. Put, as above, $m^* = |\widetilde{P}|$, and note that
$m_0 + m^* = m$.

Handling points and curves on $Z(f)$ is done as in the proof of Theorem 1.4~in
\cite{SharirSolomon} and in the two preceding sections. 
We recall that it considers each irreducible component of
$Z(f)$ separately, and distinguishes between the case where the component is 
infinitely ruled by curves of $\C$, and the case where it is not. 

We apply a variant of the inductive argument used in 
\cite[Proof of Theorem 1.4]{SharirSolomon}; see also the preceding sections. 
Briefly, the analysis in \cite{SharirSolomon} handles each irreducible component 
of $Z(f)$ separately. Enumerate these components as $Z(f_1), \ldots, Z(f_k)$, 
for suitable irreducible polynomials $f_1, \ldots, f_k$, of respective degrees
$D_1, \ldots, D_k$, where $\sum_{i=1}^k D_i \le D$. Consider first the case of 
components that are not infinitely ruled by curves of $\C$, and for simplicity 
keep the above notations for just these components. Let $P_i$ and $C_i$ denote 
the sets of all points of $\widetilde{P}$ (resp., curves of $\widetilde{C}$) 
that are contained in $Z(f_i)$; note that if a curve is contained in $Z(f)$ it 
must belong to at least one $C_i$. More precisely, we assign every point $p$ of
$\widetilde{P}$ (resp., curve $c$ of $\widetilde{C}$) to the first component
$Z(f_i)$, in the above order, that contains $p$ (resp., $c$). Let $P_i$ (resp., $C_i$) 
denote the set of points of $\widetilde{P}$ (resp., curves of $\widetilde{C}$) that are
assigned to $Z(f_i)$, for $i = 1, \ldots, k$.
 
The number of `cross-components' incidences, between points and curves
assigned to different components, can be bounded by $O(nD)$, as they occur 
when a curve crosses a component that does not fully contain it\footnote{%
  The assignment of points and curves to the components of $Z(f)$ ensures that 
  this indeed must be the case.
},
and the bound $O(nD)$ is asymptotically subsumed by the bound in
(\ref{eq:gen_in_cells_final}). We therefore focus on
bounding the number of incidences between points
and curves that are assigned to the same component $Z(f_i)$, over all $i$. 

By Theorem~\ref{thm:inf_ruled_exceptional}, there exist absolute constants $c$, $t$, such 
that there are at most $c D_i^2$ ``exceptional'' curves in $C_i$, namely, curves that 
contain at least $c D_i$ $t$-rich points of $P_i$. Denote, as before, the number 
of $t$-rich (resp., $t$-poor) points as $m_{rich}$ (resp., $m_{poor}$), 
so $m_{rich} + m_{poor} = m^*$. By choosing $a$ and $a'$
(in the definitions of $D$) sufficiently small, we can ensure that
$\sum_i D_i^2 \le (\sum_i D_i)^2 \le D^2 \le n/(2c)$.

Arguing as before, the number of incidences on the non-exceptional curves with 
the points of $P_i$, summed over all components $Z(f_i)$, is $O(m_{poor} + nD)$. 

For the exceptional curves, we simply apply induction, as their overall number, is at most
$c \sum_i D_i^2 \le c D^2 \le n/2$. Note that in this inductive step we only need to 
consider the $t$-rich points, as the $t$-poor points have already been taken care of.

Pruning away the $t$-poor points, we put $m_i = |P_i|$, and $m' := \sum_i m_i$; 
this is the number of ($t$-rich) points assigned to components that are not 
infinitely ruled. By the induction hypothesis, based on the bound in
(\ref{eq:gen_induction}), the incidence bound between the points and curves 
assigned to $Z(f_i)$ is at most
\begin{equation} \label{eq:gen_intermediate_bound}
A \left( m_i^{3/5} (c D_i^2)^{3/5} + 
  (m_i^{11/15}(c D_i^2)^{2/5} + (c D_i^2)^{8/9})\delta^{1/3} + 
   m_i^{2/3}(c D_i^2)^{1/3}\pi^{1/3} + m_i + c D_i^2 \right) .
\end{equation}

(\ref{eq:gen_intermediate_bound}), the number of other incidences on $Z(f)$ is
$O(m_{poor} + nD)$.

We now sum (\ref{eq:gen_intermediate_bound}) over $i$. The sum of the fifth 
terms is bounded by $m'$, since $\sum_i m_i = m'$. The sum of the sixth terms is
bounded by $n / 2$, as already shown. For the rest of the terms, 
first note that for any exponent $\alpha \ge 1$ the following holds:
\begin{equation} \label{gen:ineq1}
  \sum_i D_i^\alpha \le 
  \left( \sum_i D_i \right)^\alpha \le D^\alpha \le \frac{ n^{\alpha/2}}{(2c)^{\alpha/2}} ,
\end{equation}
This immediately means that the sum of the third terms is bounded by
$c^{8/9}\delta^{1/3} \sum_i D_i^{16/9} \le n^{8/9}\delta^{1/3}/2^{8/9}$.
Now apply Hölder's inequality for the sum of the first terms:
\begin{equation*}
  \sum_i m_i^{3/5}(c D_i^2)^{3/5} = c^{3/5} \sum_i m_i^{3/5}D_i^{6/5} \le
  c^{3/5} \left( \sum_i m_i \right)^{3/5} \left( \sum_i D_i^3 \right)^{2/5} =
  \frac{m^{3/5}n^{3/5}}{2^{3/5}},
\end{equation*}
and for the sum of the second terms:
\begin{align*}
  & \sum_i m_i^{11/15}(c D_i^2)^{2/5}\delta^{1/3} = 
  c^{2/5}\delta^{1/3} \sum_i m_i^{11/15}D_i^{4/5} \\ & \le 
  c^{2/5}\delta^{1/3} \left( \sum_i m_i \right)^{11/15} \left( \sum_i D_i^3 \right)^{4/15} \le
  \frac{\delta^{1/3}m^{11/15}n^{2/5}}{2^{2/5}}.
\end{align*}
For the sum of the fourth terms, we use Hölder's inequality again, and get: 
\begin{equation*}
  \sum_i m_i^{2/3}(c D_i^2)^{1/3}\pi^{1/3} \le
  \left( \sum_i m_i \right)^{2/3} \left( \sum_i c D_i^2 \right)^{1/3} \pi^{1/3} \le
  \frac{m^{2/3}n^{1/3}\pi^{1/3}}{2^{1/3}} . 
\end{equation*}
Finally, combining the bounds for each of these sums, the overall number of incidences 
between points and curves assigned to the same (not infinitely ruled) component of $Z(f)$
is at most 
\small
\begin{equation} \label{eq:gen_total_sum}
  A \left(
    \frac{m^{3/5}n^{3/5}}{2^{3/5}} + 
    \left( \frac{m^{11/15}n^{2/5}}{2^{2/5}} + \frac{n^{8/9}}{2^{8/9}} \right)\delta^{1/3} +
    \frac{m^{2/3}n^{1/3}\pi^{1/3}}{2^{1/3}} 
    + m' + n / 2
  \right)  ,
\end{equation}
\normalsize
to which we add the bound $O(m_{poor} + nD)$ for the number of incidences 
between points and curves assigned to different components, as well as for the number of 
incidences on non-exceptional curves over all (not infinitely ruled) components.

Consider next the components of $Z(f)$ that are infinitely ruled by curves of
$\C$. Let $m''$ denote the number of points of $P$ assigned to these components,
so $m' + m'' = m^*$. Again, arguing as in the proof of 
Theorem 1.4~in \cite{SharirSolomon}, using the fact that the curves of $\C$
have two degrees of freedom, the number of incidences between points and curves 
assigned to the same (infinitely ruled) component $Z(f_i)$ is
$O(m_i^{2/3}n_i^{1/3}\pi^{1/3} + m_i + n_i)$, where $m_i$ is as defined above, 
$n_i = |C_i|$, and $\pi$ is the maximum number of curves of $C$ on any surface 
that is infinitely ruled by curves of $\C$. Summing over $i$ and using Hölder's 
inequality, the total number of these incidences is
$O(m^{2/3}n^{1/3}\pi^{1/3} + m'' + n)$. The number of incidences between 
points and curves assigned to different components is
$O(nD)$, as noted earlier, which is asymptotically subsumed in the overall 
bound. 

We now add up all the sub-bounds obtained so far, within the cells and on $Z(f)$. We replace
the bound in (\ref{eq:gen_total_sum}) by (the larger bound)
\[
A/2^{1/3} \left(
  m^{3/5}n^{3/5} + (m^{11/15}n^{2/5} + n^{8/9})\delta^{1/3} + m^{2/3}n^{1/3}\pi^{1/3} + n
\right) + A m' .
\]
The sum of all the bounds is then at most
\begin{align*}
& A/2^{1/3} \left(
  m^{3/5}n^{3/5} + (m^{11/15}n^{2/5} + n^{8/9})\delta^{1/3} + m^{2/3}n^{1/3}\pi^{1/3} + n
\right) + A m'  \\
& + B_1 \left(m^{3/5}n^{3/5} + (m^{11/15}n^{2/5} + n^{8/9})\delta^{1/3} + m_0 + n \right) \\
& + B_2 \left(m^{2/3}n^{1/3}\pi^{1/3} + m'' + n \right) \\
& + B_3 \left(m_{poor} + m^{3/5}n^{3/5} + n \right)
,
\end{align*}
for suitable constants $B_1, B_2, B_3$. It follows that if we choose $A$ 
sufficiently large, the overall bound is at most
\begin{equation*}
  I(P,C) \le A\left( m^{3/5}n^{3/5} + (m^{11/15}n^{2/5} + n^{8/9})\delta^{1/3} + m^{2/3}n^{1/3}\pi^{1/3} + m + n \right) .
\end{equation*}
This establishes the induction step and thus completes the proof.
$\Box$\medskip

The bound in (\ref{eq:gen_induction}) can be sharpened to 
the ``ideal'' bound $O(m^{3/5}n^{3/5} + m + n)$ when $\delta$ and $\pi$ are not 
too large. Specifically, for the range $m < n^{2/3}$ (where the term
$n^{8/9}\delta^{1/3}$ is the dominant $\delta$-dependent term) we require
$\delta < n^{1/3}$. For the range $m \ge n^{2/3}$ we require that
$\delta < \frac{n^{3/5}}{m^{2/5}}$, which is also at most $n^{1/3}$. For the
$\pi$-term, we want it to be dominated by $m^{3/5}n^{3/5} + m$. That is, we want
to have $m^{2/3}n^{1/3}\pi^{1/3} \le m^{3/5}n^{3/5} + m$, or
$\pi \le \frac{n^{4/5}}{m^{1/5}} + \frac{m}{n}$. We thus require that 
\[
\pi \le 
\begin{cases}
\frac{n^{4/5}}{m^{1/5}}  ,\quad\text{when}\quad m < n^{3/2} \\
\frac{m}{n} ,\quad\text{otherwise}.
\end{cases}
\]
It is easily checked that when $\pi \le n^{1/2}$ the inequality holds in both 
cases. We have thus shown:
\begin{theorem} \label{thm:gen_final}
  Let $C$ be a set of $n$ curves in $\reals^3$ that are taken from a $3$-parameterizable
  family $\C$ with almost two degrees of freedom, and let $P$ be a set of $m$ points in
  $\reals^3$ whose duals are all curves. Assume that no surface that is infinitely ruled 
  by the curves of $\C$ contains more than $\pi$ curves from $C$. Let $\P^*$ be the
  family of curves in dual $3$-space that are dual to the (suitable subset of) points of
  $\reals^3$, with respect to the curves of $\C$, and assume that no surface that is 
  infinitely ruled by curves of $\P^*$ contains more than $\delta$ curves dual to the 
  points of $P$, and that not all pairs of curves of $\C$ intersect. Then
  \[
  I(P, C) = O\left(
  m^{3/5}n^{3/5} + (m^{11/15}n^{2/5} + n^{8/9})\delta^{1/3} +
  m^{2/3}n^{1/3}\pi^{1/3} + m + n \right) .
  \]
  If $\pi = O(n^{1/2})$ and $\delta = O(n^{1/3})$ then the bound becomes
  \[
  I(P, C) = O(m^{3/5}n^{3/5} + m + n).
  \]
\end{theorem}

\subsection{Lines in three dimensions} \label{subsec:gen-lines}

We conclude this section by considering the special case of lines in $\reals^3$ that have 
almost two degrees of freedom and are $3$-parameterizable. This case, a variant of which  
was also considered by Guth et al., will be used to illustrate the general machinery 
developed in this section.

So let $\L$ be a family of lines in $\reals^3$ that have almost two degrees of freedom, and
are $3$-parameterizable, let $L$ be a finite subset of $n$ lines from $\L$, and let $P$ be 
a set of $m$ points in $\reals^3$ (whose duals, with respect to $\L$, are curves). We recall
that the only surfaces that are infinitely ruled by lines are planes.

We apply the general machinery presented in the previous subsections. To apply 
Theorem~\ref{thm:gen_final}, we examine its assumptions in the context of such lines. 
We note that if every pair of lines in $\L$ 
intersect then either all the lines of $\L$ are coplanar or all are concurrent. We can rule
out both cases, since in either of them the lines are only $2$-parameterizable. In the primal
setup we have points and lines, but the dual setup depends on the specific parameterization
of the lines. For example, in the case where $\L$ consists of horizontal lines, each line
$\ell \in \L$ has a parameterization of the form $y = ax + b$, $z = c$, and is represented in 
the dual as the point $\ell^* = (a, b, c)$. The dual $p^*$ of a point $p = (\xi, \eta, \zeta)$ 
is the set of all dual points $(a, b, c)$ such that $\eta = a\xi + b$, $\zeta = c$, which is 
again a horizontal line. In general, the structure of the dual curves (when they are indeed 
curves) may be more complex.

We can thus apply the preceding analysis, and obtain:

\begin{theorem} \label{thm:lines:gs}
Let $\L$ be a family of lines in $\reals^3$ that has almost two degrees of freedom and is
$3$-parameterizable. Let $L$ be a set of $n$ lines from $\L$, and let $P$ be a set of $m$ 
points in $\reals^3$, all of whose duals are curves. Assume that no plane contains more than
$\pi$ lines of $L$ and that no dual surface, that is infinitely ruled by dual curves, contains
more than $\delta$ curves that are dual to the points of $P$. Then we have
\[
I(P,L) = O\left(
  m^{3/5}n^{3/5} + (m^{11/15}n^{2/5} + n^{8/9})\delta^{1/3} + m^{2/3}n^{1/3}\pi^{1/3} + m + n
\right) .
\]
In particular, if $\pi = O(n^{1/2})$ and $\delta = O(n^{1/3})$ then
\[
I(P, C) = O\left( m^{3/5}n^{3/5} + m + n \right) .
\]
\end{theorem}


\section{Conclusion} \label{ch:conclusion}

The elegant bound $O(m^{3/5}n^{3/5} + m + n)$ on the number of incidences, derived in
Theorems~\ref{thm:anchored_main} and \ref{thm:tangent_main} (and in the special
cases of Theorems~\ref{thm:gen_final} and \ref{thm:lines:gs}), improves upon
the best bounds for a family of curves with standard two degrees of freedom.  
Comparing this bound with the more cumbersome-looking general bound in
Theorem~\ref{thm:gen_final}, indicates that a major step in extending 
the technique of this paper to other instances of the problem, is
analyzing the structure, or establishing the nonexistence, of surfaces
that are infinitely ruled by the given curves or by the dual curves. 
This seems to be a rich area for further research, which calls for sophisticated
tools from algebraic geometry.

Specific subproblems that are still not resolved, in their full generality, are:
(a) Understand and characterize the existence of dual curves. (b) As just
mentioned, understand and characterize the existence of surfaces that are 
infinitely ruled by the family of curves, as well as of dual surfaces 
that are infinitely ruled by the family of dual curves. (c) Obtain improved
bounds, if at all possible, for the number of incidences between points and 
curves that lie on such a surface, both in the primal and in the dual setups.

In particular, it would be interesting to investigate whether ideas similar to those used in 
distinguishing between rich and poor points, given in 
Section~\ref{subsec:tangent-coplanar_lines}, can be developed to
reduce the threshold on the number of primal or dual curves that lie on a surface that is
infinitely ruled by such curves.

A natural open problem, which we have yet to make progress on, is to generalize the 
bounds and techniques from this paper to families of curves in three dimension with almost
$s$ degrees of freedom, for larger constants $s \ge 3$. For instance, the problem of 
bounding the number of incidences between (non-anchored) unit circles and points in three 
dimensions falls under this general setup for $s = 3$, since unit circles (in any dimension)
have almost three degrees of freedom. A specific goal here is to improve the bound
(\ref{eq:ssz:unit}) of \cite{SSZ} for non-anchored unit circles. 

A simple, albeit unsatisfactory, way of handling the case $s \ge 3$ is to use anchoring. For example, for the case
of unit circles in $\reals^3$, we fix a point $p_0$ of $P$, consider the subfamily $\C_{p_0}$
of the unit circles that are incident to $p_0$, apply the bound obtained in
Theorem~\ref{thm:anchored_main} to $P$ and the set $C_{p_0}$ of the circles of $C$ that are
incident to $p_0$, and then combine these bounds, over all $p_0 \in P$, to obtain the 
desired bound. We believe that this coarse (and weak) approach can be considerably 
improved by a direct approach that treats all the circles of $C$ together, and leave 
this as yet another interesting open problem for further research.

A final open question is whether the bound $O(m^{3/5}n^{3/5} + m + n)$ is tight, for any
instance of the setup considered in this paper. We strongly suspect that the bound is not
tight.

\paragraph{Acknowledgements.}
The authors would like to thank the
anonymous reviewers of the earlier conference version of this paper for their helpful comments.

\bibliographystyle{apalike}
\bibliography{references} 

\end{document}

%% file: figures/directed_points_intro.tikz
\begin{tikzpicture}[line cap=round,line join=round,>=triangle 45,x=1.6666666666666667cm,y=1.6666666666666667cm]
\clip(0.,0.) rectangle (6.,6.);
\draw [line width=1.2pt,domain=0.:6.] plot(\x,{(--1.815687992976965--1.2536745554665352*\x)/1.049587999925472});
\draw [->,line width=1.2pt] (1.1625460944296653,3.118502165947827) -- (2.579907059670836,4.811461096652557);
\draw [line width=2.pt] (2.4162206498962004,2.068914166022355) circle (2.7250559936203183cm);
\begin{scriptsize}
\draw [fill=black] (1.1625460944296653,3.118502165947827) circle (3.0pt);
\draw[color=black] (1.1771237055397412,3.6068521381353733) node {\Large $p$};
\draw[color=black] (1.9934699277039967,4.714750582501149) node {\Large $u$};
\draw[color=black] (2.795238538758176,3.329877527043929) node {\Large $c$};
\end{scriptsize}
\end{tikzpicture}

%% file: figures/anchored_circles.tikz
\begin{tikzpicture}[line cap=round,line join=round,>=triangle 45,x=1.6666666666666667cm,y=1.6666666666666667cm]
\clip(0.,0.) rectangle (6.,6.);
\draw [->,line width=1.2pt] (2.3098730887550647,2.339404971120432) -- (2.182381125325167,5.33571437516819);
\draw [->,line width=1.2pt] (2.3098730887550647,2.339404971120432) -- (4.883681211120088,2.3284525961316462);
\draw [->,line width=1.2pt] (2.3098730887550647,2.339404971120432) -- (0.682381125325167,0.8357143751681897);
\draw [line width=1.2pt] (2.3098730887550647,2.339404971120432) circle (3.75cm);
\draw [rotate around={42.77407610488608:(3.961456805817878,3.8674007909880728)},line width=1.2pt,dash pattern=on 3pt off 3pt] (3.961456805817878,3.8674007909880728) ellipse (3.75cm and 1.3621953859084281cm);
\draw [rotate around={-0.24381035610405302:(2.309873088755064,2.3394049711204326)},line width=1.2pt,dotted] (2.309873088755064,2.3394049711204326) ellipse (3.75cm and 1.5050934761733104cm);
\begin{scriptsize}
\draw [fill=black] (2.3098730887550647,2.339404971120432) circle (3.0pt);
\draw[color=black] (2.2824921512830962,2.098452721367109) node {\Large $o$};
\draw[color=black] (2.0196351515521997,4.803689343597591) node {\Large $z$};
\draw[color=black] (4.100586399421794,2.186071721277408) node {\Large $y$};
\draw[color=black] (1.1982070273931493,0.8936914726004975) node {\Large $x$};
\draw[color=black] (0.48082646562757847,4.458689531450789) node {\Large $\sph(o, 1)$};
\draw [fill=black] (3.9614568058178765,3.867400790988073) circle (3.0pt);
\draw[color=black] (3.0655869629813908,3.8672612820562704) node {\Large $c$};
\end{scriptsize}
\end{tikzpicture}

%% file: figures/infruled_lemma.tikz
\begin{tikzpicture}[line cap=round,line join=round,>=triangle 45,x=1.6666666666666667cm,y=1.6666666666666667cm]
\clip(0.,0.) rectangle (6.,6.);
\draw [rotate around={90.:(2.2551577916179464,2.813460773905509)},line width=1.2pt] (2.2551577916179464,2.813460773905509) ellipse (3.725360227913895cm and 2.4970545756268243cm);
\draw [rotate around={166.7489285396966:(4.006214636834334,4.628882101648077)},line width=1.2pt] (4.006214636834334,4.628882101648077) ellipse (3.001311688413949cm and 1.0151379733798214cm);
\draw [rotate around={164.10052126403158:(3.853556820602504,4.496430285668469)},line width=1.2pt,dash pattern=on 3pt off 3pt] (3.853556820602504,4.496430285668469) ellipse (2.824161607481004cm and 1.8546550645363507cm);
\begin{scriptsize}
\draw [fill=black] (2.2551577916179464,5.048676910653849) circle (3.0pt);
\draw[color=black] (1.9288873562747777,5.394523572117609) node {\Large $O$};
\draw[color=black] (0.46719580593738036,1.518430800240753) node {\Large $\gamma$};
\draw [fill=black] (3.4557792069868203,4.150532416626936) circle (3.0pt);
\draw[color=black] (3.5471887155768957,4.415712266088099) node {\Large $p$};
\draw [fill=black] (5.465658875394729,3.9785098827282526) circle (3.0pt);
\draw[color=black] (5.7788784933241715,3.9328320217802086) node {\Large $r$};
\draw[color=black] (4.251932855918141,4.911643327809718) node {\Large $\gamma'$};
\draw[color=black] (4.760914735053484,3.2019862466115083) node {\Large $\gamma''$};
\end{scriptsize}
\end{tikzpicture}

%% file: figures/directed_points.tikz
\begin{tikzpicture}[line cap=round,line join=round,>=triangle 45,x=1.6666666666666667cm,y=1.6666666666666667cm]
\clip(0.,0.) rectangle (6.,6.);
\draw [line width=1.2pt,domain=0.:6.] plot(\x,{(-1.851194296119771--1.1662088888060802*\x)/0.8309238332743316});
\draw [line width=1.2pt,domain=0.:6.] plot(\x,{(--9.85253794255135-0.3061298333115957*\x)/2.1720640554013246});
\draw [line width=1.2pt,dash pattern=on 3pt off 3pt] (1.1625460944296646,4.372176721414363)-- (1.0042417062805025,3.2489693959750703);
\draw [line width=1.2pt,dash pattern=on 3pt off 3pt] (2.926437038748859,1.8794052215913666)-- (1.0042417062805025,3.2489693959750703);
\draw [->,line width=1.2pt] (1.1625460944296646,4.372176721414363) -- (2.5405913269992255,4.177955581253553);
\draw [->,line width=1.2pt] (2.926437038748859,1.8794052215913666) -- (3.5067931399417356,2.693940100458562);
\begin{scriptsize}
\draw [fill=black] (1.1625460944296646,4.372176721414363) circle (3.0pt);
\draw[color=black] (1.1771237055397408,4.860526693601909) node {\Large $p$};
\draw [fill=black] (2.926437038748859,1.8794052215913666) circle (3.0pt);
\draw[color=black] (3.101368372069771,1.7117626938254926) node {\Large $q$};
\draw [fill=black] (1.0042417062805025,3.2489693959750703) circle (3.0pt);
\draw[color=black] (0.681484927797157,3.009170082622257) node {\Large $w$};
\draw[color=black] (2.4453758721163514,4.554396860290313) node {\Large $u$};
\draw[color=black] (3.6115847609224305,2.513531304879673) node {\Large $v$};
\end{scriptsize}
\end{tikzpicture}

%% file: figures/infruled.tikz
\begin{tikzpicture}[line cap=round,line join=round,>=triangle 45,x=1.6666666666666667cm,y=1.6666666666666667cm]
\clip(0.,0.) rectangle (6.,6.);
\draw [line width=1.2pt,domain=0.:6.] plot(\x,{(--0.35629069542589775-3.0467207220058845*\x)/-1.5598043887781314});
\draw [line width=1.2pt] (1.3532650361115104,4.388436588317695) circle (1.4989584494012045cm);
\draw [line width=1.2pt] (3.5521209056411687,3.2627065498503587) circle (2.618158683332134cm);
\draw [line width=1.2pt] (1.6092602017774782,4.257376862354735) circle (1.0196354991954182cm);
\draw [line width=1.2pt] (3.8186053482588433,3.126276715591549) circle (3.1171216574459426cm);
\draw [line width=1.2pt] (1.2132805705058343,4.460103276450744) circle (1.7610640515227887cm);
\draw [line width=1.2pt] (2.1538236499148335,3.9785812214423095)-- (0.04557416382908208,-0.139401419603693);
\begin{scriptsize}
\draw [fill=black] (2.1538236499148335,3.9785812214423095) circle (3.0pt);
\draw[color=black] (2.8608377887535195,3.8473827214516256) node {\Large $(x_0, y_0)$};
\draw[color=black] (1.359343844415692,1.048481388317033) node {\Large $\ell_{x_0, y_0, z_0}$};
\end{scriptsize}
\end{tikzpicture}

%% file: figures/fig012.tikz
\begin{tikzpicture}[line cap=round,line join=round,>=triangle 45,x=1.6666666666666667cm,y=1.6666666666666667cm]
\clip(0.,0.) rectangle (6.,6.);
\draw [line width=1.2pt] (0.23449973145063246,3.630053057781741)-- (0.976699717769274,0.503338221801079);
\draw [line width=1.2pt] (5.587814526387217,1.6719084130261748)-- (2.934844362524839,0.47175524365986027);
\draw [line width=1.2pt] (0.976699717769274,5.461865789972432)-- (5.4772741028929515,4.640708358300743);
\draw [line width=1.2pt] (3.0854560588159297,5.0771102470446206)-- (2.520843671358557,1.9826000465570977);
\draw [line width=1.2pt] (4.111835710739724,1.0042037107094512)-- (2.499253434582938,4.568859268529714);
\draw [line width=1.2pt] (0.6329439484035574,1.9515008246609071)-- (4.280578280331767,2.8173534186034614);
\begin{scriptsize}
\draw [fill=black] (0.6329439484035574,1.9515008246609071) circle (3.0pt);
\draw[color=black] (2.074208208176627,1.6087424567437372) node {\Large $(p_0=(x_0,y_0), z_0)$};
\draw [fill=black] (3.0854560588159297,5.0771102470446206) circle (3.0pt);
\draw[color=black] (4.07972732014402,5.477657279043041) node {\Large $(p_2=(x_2,y_2), z_2)$};
\draw [fill=black] (4.111835710739724,1.0042037107094512) circle (3.0pt);
\draw[color=black] (4.506097525050474,0.4401722655186414) node {\Large $(p_1=(x_1,y_1), z_1)$};
\draw [fill=black] (2.6003796498392524,2.418518390153218) circle (3.0pt);
\draw[color=black] (2.295289055165158,2.8878530714630988) node {\Large $w_{02}$};
\draw [fill=black] (2.850896445959293,3.7915431381188243) circle (3.0pt);
\draw[color=black] (3.2901528666135507,3.9300913501233197) node {\Large $w_{12}$};
\draw [fill=black] (3.387502766179883,2.6053607460522565) circle (3.0pt);
\draw[color=black] (3.574399669884519,3.0299764730985834) node {\Large $w_{01}$};
\end{scriptsize}
\end{tikzpicture}

%% file: figures/power.tikz
\begin{tikzpicture}[line cap=round,line join=round,>=triangle 45,x=1.6666666666666667cm,y=1.6666666666666667cm]
\clip(0.,0.) rectangle (6.,6.);
\draw [line width=1.2pt,domain=0.:6.] plot(\x,{(--1.9092899928305709-3.046720722005884*\x)/-1.5598043887781314});
\draw [line width=1.2pt] (1.5719292027626501,3.8199097550247307) circle (1.4989584494012047cm);
\draw [line width=1.2pt] (3.7707850722923073,2.6941797165573944) circle (2.6181586833321324cm);
\draw [line width=1.2pt] (1.8279243684286177,3.688850029061771) circle (1.0196354991954186cm);
\draw [line width=1.2pt] (4.037269514909982,2.5577498822985847) circle (3.117121657445942cm);
\draw [line width=1.2pt] (1.431944737156974,3.89157644315778) circle (1.7610640515227882cm);
\draw [->,line width=1.2pt] (2.372487816565973,3.4100543881493453) -- (0.8038842410490425,0.34614646961627393);
\begin{scriptsize}
\draw [fill=black] (2.372487816565973,3.4100543881493453) circle (3.0pt);
\draw[color=black] (2.838971372088405,3.2424118603834713) node {\Large $p$};
\draw[color=black] (1.490542344406376,1.1213694438674129) node {\Large $\ell_{p,u}$};
\draw [fill=black] (3.046686597269657,4.72694733363598) circle (3.0pt);
\draw[color=black] (2.824393760978329,5.1958117491336555) node {\Large $w$};
\end{scriptsize}
\end{tikzpicture}

%% file: figures/power2.tikz
\begin{tikzpicture}[line cap=round,line join=round,>=triangle 45,x=1.6666666666666667cm,y=1.6666666666666667cm]
\clip(0.,0.) rectangle (6.,6.);
\draw [line width=1.2pt,domain=0.:6.] plot(\x,{(--4.950237014557289--0.789574453530471*\x)/2.084476557320439});
\draw [->,line width=1.2pt] (1.229363542899024,2.840478604251278) -- (3.313840100219463,3.630053057781749);
\draw [line width=1.2pt] (3.313840100219463,3.630053057781749) circle (3.7150109674155494cm);
\draw [line width=1.2pt,dash pattern=on 3pt off 3pt] (1.4766504626957588,2.187641135987899) circle (1.1635047398450646cm);
\draw [line width=1.2pt,dash pattern=on 3pt off 3pt] (1.6995257250253804,1.5992504434376997) circle (2.2121506784490137cm);
\draw [line width=1.2pt,dash pattern=on 3pt off 3pt] (1.6333447705284978,1.7739681633094693) circle (1.9007639932661202cm);
\draw [line width=1.2pt,dash pattern=on 3pt off 3pt] (3.313840100219463,3.630053057781749)-- (5.228030259917664,2.487973214600464);
\begin{scriptsize}
\draw [fill=black] (1.229363542899024,2.840478604251278) circle (3.0pt);
\draw[color=black] (0.8187848270631799,3.0378722176338924) node {\Large $p$};
\draw [fill=black] (3.313840100219463,3.630053057781749) circle (3.0pt);
\draw[color=black] (3.4243805237137295,3.953778583729239) node {\Large $w$};
\draw[color=black] (2.0978954417825406,3.6063658241758314) node {\Large $u$};
\draw[color=black] (2.2400188434180253,5.343429621942867) node {\Large $\gamma$};
\draw[color=black] (4.435035824232731,3.306327531834252) node {\Large $\sqrt{\rho}$};
\end{scriptsize}
\end{tikzpicture}

%% file: figures/unique1.tikz
\begin{tikzpicture}[line cap=round,line join=round,>=triangle 45,x=1.6666666666666667cm,y=1.6666666666666667cm]
\clip(0.,0.) rectangle (6.,6.);
\draw [shift={(5.271984744975027,2.840478604251278)},line width=2.pt,fill=black,fill opacity=0.10000000149011612] (0,0) -- (160.39376499911467:0.34352561161934525) arc (160.39376499911467:205.50693808884898:0.34352561161934525) -- cycle;
\draw[line width=2.pt,fill=black,fill opacity=0.10000000149011612] (2.0437692785085018,2.380303281134107) -- (2.094374136143839,2.6178828915869614) -- (1.8567945256909852,2.6684877492222983) -- (1.806189668055648,2.4309081387694444) -- cycle; 
\draw [line width=1.2pt] (2.131691133685596,3.959071936076426) circle (2.6040757560275374cm);
\draw [->,line width=1.2pt] (5.271984744975027,2.840478604251278) -- (4.119206042820542,5.25657643205452);
\draw [line width=1.2pt,domain=0.:6.] plot(\x,{(-0.785442453562541-1.1527787021544853*\x)/-2.416097827803242});
\draw [line width=1.2pt] (3.608671252128911,2.046871512893329) circle (3.0715649245844863cm);
\draw [line width=1.2pt] (2.131691133685596,3.959071936076426)-- (5.271984744975027,2.840478604251278);
\draw [line width=1.2pt] (3.608671252128911,2.046871512893329)-- (2.131691133685596,3.959071936076426);
\draw [line width=0.8pt] (3.608671252128911,2.046871512893329)-- (5.271984744975027,2.840478604251278);
\draw [line width=1.2pt,dash pattern=on 3pt off 3pt] (2.131691133685596,3.959071936076426)-- (1.806189668055648,2.4309081387694444);
\draw [line width=1.2pt,dash pattern=on 3pt off 3pt] (1.806189668055648,2.4309081387694444)-- (3.608671252128911,2.046871512893329);
\begin{scriptsize}
\draw [fill=black] (2.131691133685596,3.959071936076426) circle (3.0pt);
\draw[color=black] (1.9942808890378578,4.193814437349646) node {\Large $w$};
\draw[color=black] (1.1125651525482052,5.464859200341226) node {\Large $\gamma$};
\draw [fill=black] (5.271984744975027,2.840478604251278) circle (3.0pt);
\draw[color=black] (5.498242127555179,2.796810283430972) node {\Large $p'$};
\draw[color=black] (4.834092611757778,4.308322974556094) node {\Large $u'$};
\draw [fill=black] (3.608671252128911,2.046871512893329) circle (3.0pt);
\draw[color=black] (3.591674983067813,1.8521148514777699) node {\Large $x$};
\draw[color=black] (4.41041102409392,2.2700710122813077) node {\Large $r$};
\draw[color=black] (4.799740050595844,2.842613698313551) node {\Large $\alpha$};
\draw [fill=black] (5.338106943725981,1.4101013977279457) circle (0.5pt);
\draw[color=black] (5.532594688717114,1.445609544394877) node {\Large $c$};
\draw [fill=black] (1.806189668055648,2.4309081387694444) circle (3.0pt);
\end{scriptsize}
\end{tikzpicture}

%% file: figures/unique21.tikz
\begin{tikzpicture}[line cap=round,line join=round,>=triangle 45,x=1.6666666666666667cm,y=1.6666666666666667cm]
\clip(0.,0.) rectangle (6.,6.);
\draw [->,line width=1.2pt] (3.1875081876545877,2.035112661650197) -- (3.0295932969484944,5.351325366478176);
\draw [->,line width=1.2pt] (5.477274102892949,2.9983934949573716) -- (0.7398273817101328,3.6616360359229674);
\draw [line width=1.2pt] (4.238090133385114,2.0851403733516505) circle (1.752954012067933cm);
\draw [line width=1.2pt] (1.6589564631871745,1.9623244842946064) circle (2.5504729916097326cm);
\begin{scriptsize}
\draw [fill=black] (3.1875081876545877,2.035112661650197) circle (3.0pt);
\draw[color=black] (2.903261384383619,1.8061360701263605) node {\Large $p'$};
\draw[color=black] (3.2980486111488534,4.601229635624227) node {\Large $u'$};
\draw [fill=black] (5.477274102892949,2.9983934949573716) circle (3.0pt);
\draw[color=black] (5.524648570104777,3.3852849771873026) node {\Large $p$};
\draw[color=black] (3.7244188160553073,3.5431998678933967) node {\Large $u$};
\end{scriptsize}
\end{tikzpicture}

%% file: figures/unique22.tikz
\begin{tikzpicture}[line cap=round,line join=round,>=triangle 45,x=1.6666666666666667cm,y=1.6666666666666667cm]
\clip(0.,0.) rectangle (6.,6.);
\draw [->,line width=1.2pt] (3.5033379690667763,1.529785011390696) -- (4.608742204009434,4.98812111785416);
\draw [line width=1.2pt,dash pattern=on 3pt off 3pt] (2.0029882195389144,2.0093488582717467) circle (2.6252147288595853cm);
\begin{scriptsize}
\draw [fill=black] (3.5033379690667763,1.529785011390696) circle (3.0pt);
\draw[color=black] (3.7875847723377456,1.5376807559260006) node {\Large $p'$};
\draw[color=black] (4.70349113843309,3.6853232695288822) node {\Large $u=u'$};
\draw [fill=black] (3.803000152656347,2.467299557192354) circle (3.0pt);
\draw[color=black] (4.056040086538105,2.6115020127274415) node {\Large $p$};
\end{scriptsize}
\end{tikzpicture}